\documentclass[a4paper]{article}
\usepackage{graphicx} 

\usepackage{amsthm}
\usepackage{float}
\usepackage[section,below]{placeins} 

\usepackage{fewerfloatpages}

\restylefloat{table}
\usepackage{amsmath}
\usepackage{amssymb}
\usepackage{amsfonts,cancel}
\usepackage[usenames]{color}
\usepackage{authblk} 
\usepackage{soul} 
\usepackage{longtable}

\newcommand{\be}{\begin{equation}}
	\newcommand{\ee}{\end{equation}}
\newcommand{\bea}{\begin{eqnarray}}
	\newcommand{\eea}{\end{eqnarray}}
\newcommand{\Kdt}{{\hbox{\tiny K}}}
\newcommand{\rovno}{\!\!\!& = &\!\!\!} 

\def \d {{\rm d}}
\def \dd {{\rm d}}

\def \pul {\textstyle{\frac{1}{2}}}

\def \H {\mathcal{H}}

\def \B {\mathcal{B}}
\def \k {m}

\def \L {\Lambda}
\def \k {\kappa'}

\newcommand{\beqn}{\begin{eqnarray}}
	\newcommand{\eeqn}{\end{eqnarray}}

\usepackage{amsmath,amsthm,amssymb,cite,enumerate}

\def \d {{\rm d}}

\begin{document}

	\title{On vacuum and charged asymptotically (A)dS black holes in quadratic gravity}

\author[1,2]{George Turner \thanks{turner.math.cas.cz}}
    
	\author[1]{Vojt\v ech Pravda\thanks{pravda.math.cas.cz}}
	
	\author[1]{Alena Pravdov\'a \thanks{pravdova.math.cas.cz}}

 \affil[1]{Institute of Mathematics of the Czech Academy of Sciences, \newline \v Zitn\' a 25, 115 67 Prague 1, Czech Republic}
 
\affil[2]{ Charles University, V~Hole\v{s}ovi\v{c}k\'ach~2, 180~00 Prague 8, Czech Republic}

\maketitle

\begin{abstract}
In this paper, we study asymptotic properties of static spherically symmetric black holes in quadratic gravity with a cosmological constant $\L$. We find that for sufficiently large values of $|\L|$ these black holes are generically asymptotically (A)dS and form a three-parameter family of black holes, with free parameters being the horizon radius, $\L$, and Bach parameter $b$. 

For smaller values of $|\L|$, fine-tuning of the Bach parameter is necessary to recover asymptotically (A)dS behaviour, resulting in two distinct two-parameter families of spherically symmetric black holes with (A)dS asymptotics.
One family is the (A)dS-Schwarzschild solution, the other is a fine-tuned asymptotically (A)dS-Schwarzschild–Bach solution.

We also generalize the above solutions to electrically charged black holes, obtaining qualitatively similar asymptotic behaviour.  This adds the electric charge as an additional free parameter of these black holes. Both fine-tuned three-parameter families are distinct from  Reissner-Nordstr\" om black holes.

\end{abstract}

\section{Introduction}

Black-hole uniqueness theorems 
imply that the Schwarzschild and  Reissner-Nordstr\" om black holes are the only static, asymptotically flat black holes in general relativity in vacuum and electrovacuum, respectively \cite{Israel67,Israel68}. 

Recently, inspired by seminal work \cite{Stelle78}, it has been shown that black-hole uniqueness fails in vacuum in quadratic gravity \cite{Luetal15,Luetal15b}. Besides the Schwarzschild black hole, which remains an exact solution of this theory, another vacuum static, spherically symmetric, and asymptotically flat solution of quadratic gravity does exist \cite{Luetal15,Luetal15b}. 
 In \cite{Luetal15,Luetal15b}, the quadratic gravity field equations have been solved numerically. In \cite{Podolskyetal18,PodSvaPraPra20}, the field equations have been considerably simplified using the conformal-to-Kundt approach. This simplification has allowed the solutions to be expressed as power series with coefficients known to all orders, thanks to a recursive relation, and thus enables analytical approaches to the study of these black holes.

 These new black holes are characterized by two parameters, the horizon radius $\bar r_h$ and $b$, the Bach parameter,  corresponding to the value of the Bach-tensor invariant on the horizon \cite{Luetal15,Luetal15b,Podolskyetal18,PodSvaPraPra20}. However, for a generic value of  $b$, the spacetime is not asymptotically flat, and fine-tuning of $b$ (separately for each horizon radius $\bar r_h$)  to suppress the growing Yukawa modes and achieve asymptotic flatness is necessary. 
Thus, effectively, in quadratic gravity, there are two one-parameter families of static, spherically symmetric, asymptotically flat black holes, Schwarzschild ($b=0$) and Schwarzschild-Bach ($b (\bar r_h)\not= 0$).

The Schwarzschild-Bach family has been generalized to the case of a non-vanishing cosmological constant $\Lambda$ in \cite{Svarcetal18,PraPraPodSva21}; however, asymptotic properties of these black-hole solutions to quadratic gravity were not studied there.
Subsequently, it has been briefly discussed in \cite{PraPraOrt23} that within a
certain range of parameters of the theory and of the solution, it is not necessary to fine-tune the Bach parameter $b$  to
obtain an asymptotically AdS spacetime, thus making the Bach parameter
a new free parameter of the AdS-Schwarzschild solution.\footnote{This observation is in agreement with previous numerical results of \cite{Luetal12} in Einstein-Weyl gravity.}

This paper has two main objectives: 

(i) Our first goal is to clarify the asymptotic properties of Schwarzschild–Bach black holes with a cosmological constant \cite{Svarcetal18,PraPraPodSva21} and to distinguish generically asymptotically (A)dS solutions from those that require fine-tuning of the Bach parameter. We will refer to the corresponding black holes as generic asymptotically (A)dS-Schwarzschild–Bach (G(A)dS–Schwarzschild–Bach) black holes and fine-tuned asymptotically (A)dS-Schwarzschild–Bach (FT(A)dS–Schwarzschild–Bach) black holes, respectively.

The black-hole solutions are expressed as power series, expanded around the horizon. In the case of vanishing $\L$, the fine-tuning of the Bach parameter is computationally demanding, but in principle straightforward.  In contrast, in the case of positive $\L$, one has a power series solution around the black-hole horizon for which the radius of convergence ends below or on the cosmological horizon, and it is necessary to overcome this limitation to study asymptotics.

(ii) The second objective is to construct an electromagnetic generalization of these black holes, and study their asymptotic behaviour as in point (i).  As a side result of this derivation, we also identify other classes of charged static spherically symmetric spacetimes in quadratic gravity, which will be studied in detail elsewhere.

Electromagnetic charge has already been added to Schwarzschild-Bach black holes in quadratic gravity with $\Lambda=0$ in  \cite{Lin2017,Wu20} using numerical methods and in \cite{PraPraTur24} using infinite power series. As in the vacuum, in the presence of an electromagnetic field, fine-tuning is necessary to achieve asymptotic flatness. In contrast with the vacuum case, the Reissner-Nordstr\" om black hole appearing in general relativity is not a solution to quadratic gravity with a coupled electromagnetic field. Instead, two new two-parameter ($\bar r_h$ and charge $q$) families of charged static spherically symmetric asymptotically flat black holes appear. In the limit of vanishing charge, one of the families reduces to the Schwarzschild black hole and the other one to  
the Schwarzschild-Bach black hole. Thus, in the $\L=0$ case, we can refer to charged Schwarzschild and charged Schwarzschild-Bach black holes in quadratic gravity, both classes being distinct from  Reissner-Nordstr\" om \cite{PraPraTur24}. In this paper, we will show that such a distinction does not hold in general for $\Lambda \not=0$.
Therefore, we will refer to these black holes as
charged asymptotically (A)dS quadratic gravity (charged (A)dS-QG) black holes. 

To construct the solution for charged static spherically symmetric black holes in quadratic gravity in the presence of  $\Lambda$,
we follow the approach taken in a series of related papers in the vacuum case \cite{Podolskyetal18,PodSvaPraPra20,Svarcetal18,PraPraPodSva21,PraPraOrt23} and charged case with vanishing $\L$ \cite{PraPraTur24}. We use the conformal-to-Kundt ansatz that allows us to simplify the field equations of quadratic gravity. Further simplification follows from assuming that the Ricci scalar $R$ is a constant. This is motivated by the (vacuum) trace no-hair theorem of \cite{Nelson2010}, \cite{Luetal15b}.  

In general, we can expand the power-series solution around any point in the spacetime. However, we focus on solutions expanded around a horizon. With a cosmological constant being present, for $\Lambda>0$, the expansion can be done either around a black-hole horizon or around a cosmological horizon. As in the vacuum case, the presence of a cosmological horizon may complicate fine-tuning.

It turns out that in both vacuum and charged cases, for sufficiently large values of $|\Lambda|$, the black holes generically exhibit (A)dS asymptotics, and fine-tuning is not needed, while for smaller values of $|\Lambda|$, fine-tuning is necessary.

In Sec. \ref{sec_bcg}, we briefly summarize the necessary background concerning the field equations of quadratic gravity and the conformal-to-Kundt approach.

In Sec. \ref{sec_FEQ}, the quadratic gravity field equations (for static spherically symmetric spacetimes) coupled to an electromagnetic field in the Kundt coordinates are given, and an initial discussion relating the properties of the metric functions in the Kundt coordinates to the asymptotically (A)dS property in the physical coordinates is presented.

In Sec. \ref{sec_expansionFE}, we expand the metric functions in the Kundt coordinates as an infinite power series, substitute them in the field equations, obtain formal power-series identities, and summarize
possible leading powers of solutions. Technical derivation is deferred to Appendix \ref{expansiont_0}.
 Some of these solutions admit extremal or non-extremal horizons. In the rest of the paper, we focus on the non-extremal case.

In Sec. \ref{sec_Schwarzschild}, we discuss the Schwarzschild-(A)dS metric in the context of this work. In particular, we express this metric in the Kundt coordinates as a power series around the horizon. We then discuss how the number of horizons depends on the coefficients in the power series.  We expect qualitatively similar behaviour for vacuum Schwarzschild-Bach black holes with a small Bach parameter.

In Sec. \ref{sec_[0,1]}, we present the infinite power series solution to quadratic gravity for charged black holes with coefficients expressed in the form of a recursive formula. 

The last two Secs. \ref{tuning_vac} and \ref{tuning_elmg}, devoted to vacuum and charged cases respectively, focus on mapping the parameter space to distinguish generically asymptotically (A)dS solutions from those requiring fine-tuning of the Bach parameter. 

It is found that for sufficiently large values of $|\L|$ ($\L>\L_+$ or $\L<\L_-$, where $\L_\pm$ are of order 1 and depend on $\bar r_h$, $b$ and $q$) the black holes are generically asymptotically (A)dS, without the need for fine-tuning. In the charged case, this leads to a 4-parameter family   $(\bar r_h,b,q,\L)$ of charged generic asymptotically (A)dS quadratic gravity (charged G(A)dS-QG) black holes. 

In contrast, for $|\L| < |\L_\pm|$, fine-tuning of the Bach parameter is needed, and in the vacuum case, we obtain two classes of black holes, (A)dS-Schwarzschild and FT(A)dS-Schwarzschild-Bach, while in the charged case, we obtain two branches of charged fine-tuned asymptotically (A)dS quadratic gravity (charged FT(A)dS-QG) black holes. 
\section{Background}
\label{sec_bcg}

\subsection{Quadratic gravity}
	\label{subsec_QG}
	In quadratic gravity, terms quadratic in curvature tensors are added to the Einstein-Hilbert action. Thus, coupled with the electromagnetic field, the action reads
	\be
	S {=\int \d^4 x\, \sqrt{-g}{\cal L}}= \int \d^4 x\, \sqrt{-g}\, \Big(
	\gamma \,(R-2\Lambda) +\beta\,R^2  - \alpha\, C_{abcd}\, C^{abcd}
	-\frac{\kappa}{2} F_{ab} F^{ab}\Big)\,,
	\label{action}
	\ee
	where ${\gamma=1/(16\pi G)}$, $G$ is the Newtonian constant (we  set $G=1=c$), $\Lambda $ is the cosmological constant, and  $\alpha$, $\beta$  are {coupling} constants of quadratic gravity.
	
	The field equations following from the action \eqref{action} take the form
	\be
	\gamma \left(R_{ab} - {\pul} R\, g_{ab}+\Lambda\,g_{ab}\right)-4 \alpha\,B_{ab}
	+2\beta\left(R_{ab}-\tfrac{1}{4}R\, g_{ab}+ g_{ab}\, \Box - \nabla_b \nabla_a\right) R = \kappa T_{ab}\,, \label{fieldeqsEW}
	\ee
	where the energy momentum tensor $T_{ab}$ reads
	$$
	T_{ab} = F_{ac} {F_{b}}^{ c} - \frac{1}{4} {g_{ab} F_{cd} F^{cd}},
	$$
	and  $B_{ab}$ is the {Bach tensor}
	\be
	B_{ab} \equiv \big( \nabla^c \nabla^d + {\pul} R^{cd} \big) C_{acbd} \ , \label{defBach}
	\ee
	 traceless, symmetric, conserved, and well-behaved under a conformal transformation $g_{ab}=\Omega^2 \tilde g_{ab}$:
	\begin{equation}
		g^{ab}B_{ab}=0 \,, \qquad B_{ab}=B_{ba} \,, \qquad
		\nabla^b B_{ab}=0
		\,, \qquad B_{ab}=\Omega^{-2}\tilde B_{ab}\,.
		\label{Bachproperties}
	\end{equation}

	Motivated by the (vacuum) trace no-hair theorem of \cite{Nelson2010}, \cite{Luetal15b}, we will further assume 
	that  Ricci scalar $R$ is constant throughout the spacetime (see also related discussion in \cite{PraPraTur24}).

	Then the trace of \eqref{fieldeqsEW}  reduces to
	\be
	R=4\Lambda ,
	\label{R=const}
	\ee
	assuming  $\gamma\neq0$. 
	The quadratic gravity field equations then simplify considerably to 
	\be
	R_{ab}-\Lambda \, g_{ab}-4k\, B_{ab} = \kappa' T_{ab},
	\label{eq:feq}
	\ee
	where 
	\be
	k\equiv\frac{\alpha}{\gamma+8\beta\Lambda}, \qquad  \kappa' \equiv \frac{\kappa}{\gamma+8\beta\Lambda} \qquad (\gamma+8\beta\Lambda\neq0).
	\label{k}
	\ee 
	
We thus achieved a substantial simplification of the field equations by assuming the constancy of the Ricci scalar. Further considerable simplification can be attained by using the conformal-to-Kundt ansatz for the metric.

	\subsection{Conformal-to-Kundt ansatz}

The conformal-to-Kundt ansatz was first used in \cite{Pravdaetal17}. It has been pointed out there that all Robinson-Trautman spacetimes are conformal to Kundt, and since the (conformally well-behaved) Bach tensor is the only correction term in the quadratic-gravity equations \eqref{eq:feq}, using the conformal-to-Kundt ansatz should lead to simplification of the field equations for Robinson-Trautman spacetimes in quadratic gravity. Static, spherically symmetric spacetimes belong to the Robinson-Trautman class, and the conformal-to-Kundt ansatz was a key ingredient in the studies \cite{Podolskyetal18,PodSvaPraPra20,Svarcetal18,PraPraPodSva21,PraPraOrt23,PraPraTur24} of spherically symmetric black holes in quadratic gravity.

	In the conformal-to-Kundt ansatz, instead of using the 
	canonical static spherically symmetric metric
	\be
	\dd s^2 = -h(\bar r)\,\dd t^2+\frac{\dd \bar r^2}{f(\bar r)}+\bar r^2 (\dd \theta^2 + \sin^2 \theta \dd \phi^2), \ \ \  \ 
	\label{physmet}
	\ee	
	one employs the conformal-to-Kundt form 
	\be
	\dd s^2 \equiv \Omega^2(r) \,\dd s^2_\Kdt = \Omega^2(r)
	\Big[\,\dd \omega^2 -2\,\dd u\,\dd r+\H(r)\,\dd u^2 \,\Big]\,.
	\label{BHmetric}
	\ee

	The conformal-to-Kundt metric \eqref{BHmetric}  admits a gauge freedom \cite{Podolskyetal20}
	\be
	r \to \lambda\,r+\upsilon\,, \qquad u \to \lambda^{-1}\,u \,, 
	\label{scalingfreedom}
	\ee
	where $\lambda$ , and $\upsilon$ are constants.

   The metric \eqref{physmet} in the physical coordinates also has a time-scaling freedom 
   \be
t\ \rightarrow\ t/\sigma\,,\quad
\sigma=\mbox{const.}\not=0\,,\label{gaugetrt}
\ee
which can be used to adjust the value of
$h$ at a given radius $\bar r$ due to the rescaling ${h \ \rightarrow\ h \sigma^2}$.

	As shown in \cite{Podolskyetal18}, the metric \eqref{physmet} can be obtained (assuming $\Omega'\neq0\neq \H$) via 
	\begin{equation}
		\bar{r} = \Omega(r)\,, \qquad t = u - \int\! \frac{\dd r}{\H(r)} \,,
		\label{to static}
	\end{equation}
	leading to
	\begin{equation}
		h = -\Omega^2\, \H , \quad f = -\left(\frac{\Omega'}{\Omega}\right)^2 \H  , 
		\label{Schwarz}  
	\end{equation}
	where a prime denotes differentiation with respect to the Kundt coordinate $r$.

The conformal-to-Kundt metric \eqref{BHmetric} is also suitable for expressing curvature invariants constructed from the Bach and Weyl tensors \cite{Podolskyetal18}, as both these tensors are conformally well-behaved
\begin{align}
B_{ab}\, B^{ab} &=  \tfrac{1}{72}\,\Omega^{-8}\,\big[(\B_1)^2 + 2(\B_1+\B_2)^2\big] \,,\label{invB}\\
C_{abcd}\, C^{abcd} &=  \tfrac{1}{3}\,\Omega^{-4}\,\big({\H}'' +2\big)^2 \,, \label{invC}
\end{align}
where the functions $\B_1(r)$ and $\B_2(r)$ are two independent components of the Bach tensor,
\bea
&& \B_1 \equiv {\H}{\H}''''\,, \label{B1}\\
&& \B_2 \equiv {\H}'{\H}'''-\tfrac{1}{2}{{\H}''}^2 +2\,. \label{B2}
\eea

	\section{The field equations in Kundt coordinates} 

\label{sec_FEQ}
    
	\subsection{Maxwell equations}

    As in \cite{PraPraTur24}, for the electromagnetic field, we assume 
	\be
	\mathbf{A} = A(\bar{r}) \dd t\,,   
	\ee
	which in the Kundt coordinates becomes
	\be
	\mathbf{A} = A(r) \dd u - \frac{A(r)}{\H(r)} \dd r \,.
	\ee
	The Maxwell equations then read
	\be
	A''=0
	\ee
	and therefore can be solved exactly in the Kundt coordinates
	\be
	A= q r
	\ee
	(the constant term can be removed by gauge transformation).
    
   In contrast, the  electromagnetic field invariant can be exactly expressed in the physical coordinates
	\be
	F_{ab} F^{ab}=- \frac{2 q^2}{\Omega^4} = - \frac{2 q^2}{{\bar r}^4} .\ee

	\subsection{Field equations of quadratic gravity}
	
	Employing  the same methods as in \cite{Svarcetal18,Podolskyetal18,Podolskyetal20,Pravdaetal21,PraPraTur24},   we arrive to the expression of the field equations \eqref{eq:feq}  in the Kundt coordinates \eqref{BHmetric}
	\begin{align}
		\Omega\Omega''-2{\Omega'}^2 = &\ \tfrac{1}{3}k\,{\H}'''' \,, \label{Eq1C}\\
		\Omega\Omega'{\H}'+3\Omega'^2{\H}+\Omega^2 -\Lambda \Omega^4
		= &\ \tfrac{1}{3}k \big({\H}'{\H}'''-{\textstyle\frac{1}{2}}{{\H}''}^2 +2 \big)+ \frac{\kappa'}{2} q^2\,. \label{Eq2C}
	\end{align}
	The trace of the field equations \eqref{eq:feq}, which follows from the two above equations, reads
	\begin{equation}
		{\H}\Omega''+{\H}'\Omega'+{\textstyle \frac{1}{6}} ({\H}''+2)\Omega =
		{\textstyle \frac{2}{3}\Lambda \,\Omega^3 } \,.
		\label{traceC}
	\end{equation}

     \subsection{{Asymptotics of well-behaved solutions}}

  Recall that for the Schwarzschild-Bach black holes with vanishing $\Lambda$, fine-tuning of the Bach parameter $b$ is necessary for achieving the asymptotic flatness. In the presence of $\Lambda$, there are continuous regions (in the parameter space of $\Lambda$, $b$, and $q$) for which fine-tuning is not necessary to achieve (A)dS asymptotics. However, elsewhere, fine-tuning is also necessary. Typically, for the well-behaved solutions, $\Omega$ has a pole for a finite value of $r$ ($\bar r=\Omega\ \rightarrow\ \infty$) and $\H$ is regular there, i.e.,
 $\H ,\ \H' ,\H'' ,\ \dots <<\Omega$.
   Eq. \eqref{Eq1C} then implies
   \be
   \Omega\Omega''-2\Omega'^2=0
   \ee
   with a solution
   \be
   \Omega\rightarrow -\frac{1}{p_1 r+p_0}\,,
   \ee
   which goes to infinity for
   $r\rightarrow -p_0/p_1$ (if needed, gauge $r\rightarrow\ \frac{r-p_0}{p_1}$ leads to $\Omega=-1/r$). Then
   eq. \eqref{Eq2C} gives
   \be
   \H \rightarrow \frac{\Lambda}{3p_1^2},
   \ee
   and 
   \bea
   \frac{f}{\bar r^2}&& \rightarrow -\frac{\Lambda}{3}\,,\\
   \frac{h}{\bar r^2}&& \rightarrow -\frac{\Lambda}{3 p_1^2}\, \qquad
   \mbox{and after a gauge transformation:\ \ } \frac{h}{\bar r^2} \rightarrow - \frac{\Lambda}{3}
   \eea
   as expected since this is the asymptotic behaviour of the metric functions $f$ and $h$ in the Schwarzschild-(A)dS case.

\section{Expanding the metric around a point in a finite distance}
\label{sec_expansionFE}

In this section, we will solve  autonomous equations \eqref{Eq1C}, \eqref{Eq2C}
using  expansions in powers of  ${\Delta \equiv r-r_0}  $  around an arbitrary finite fixed value ${r_0}$,
\begin{eqnarray}
	\Omega(r) \rovno \Delta^n   \sum_{i=0}^\infty a_i \,\Delta^{i}\,, \label{rozvojomeg0}\\
	\H(r)     \rovno \Delta^p \,\sum_{i=0}^\infty c_i \,\Delta^{i}\,, \label{rozvojcalH0}
\end{eqnarray}
where $n, p$ are in principle real numbers; however, we will see that in many cases, the indicial equations restrict their possible values to integers. Naturally, we also assume ${a_0\not=0},\ {c_0\not=0}$.
Note that integer steps in  ${\Delta=r-r_0}$ in the Kundt coordinates might not be so in the physical coordinate $\bar r$; indeed, in some cases, they correspond to 
series in rational powers of $\bar\Delta=\bar r-\bar r_0$
 (see Sec. \ref{sec_specialq}).

Substituting the series \eqref{rozvojomeg0}, \eqref{rozvojcalH0}
into the field equations (\ref{Eq1C}) and (\ref{Eq2C}) and the trace equation
(\ref{traceC}) yields
\begin{align}
	&\sum_{l=2n-2}^{\infty}\Delta^{l}\sum^{l-2n+2}_{i=0}a_i\, a_{l-i-2n+2}\,(l-i-n+2)(l-3i-3n+1) \nonumber \\
	& \hspace{45.0mm}=\tfrac{1}{3}k \sum^{\infty}_{l=p-4}\Delta^{l}\,c_{l-p+4}\,(l+4)(l+3)(l+2)(l+1) \,,
	\label{KeyEq1C}
\end{align}
\begin{align}
	&\sum_{l=2n+p-2}^{\infty}\Delta^{l}\sum^{l-2n-p+2}_{j=0}\sum^{j}_{i=0}a_i\,a_{j-i}\,c_{l-j-2n-p+2}\,(j-i+n)(l-j+3i+n+2) \nonumber \\
	& \hspace{10.0mm} +\sum_{l=2n}^{\infty}\Delta^{l}\sum^{l-2n}_{i=0}a_i\,a_{l-i-2n}-\Lambda \sum_{l=4n}^{\infty}
	\Delta^{l}\sum^{l-4n}_{m=0}\bigg(\sum^{m}_{i=0}a_i\,a_{m-i}\bigg)\bigg(\sum^{l-m-4n}_{j=0}a_j\,a_{l-m-j-4n}\bigg) \nonumber \\
	& = \tfrac{1}{3}k \bigg[2+\sum^{\infty}_{l=2p-4}\Delta^{l}\sum^{l-2p+4}_{i=0}c_{i}\,c_{l-i-2p+4}\,(i+p)(l-i-p+4)
	(l-i-p+3)(l-\tfrac{3}{2}i-\tfrac{3}{2}p+\tfrac{5}{2})\bigg]
	+\frac{\kappa' q^2}{2}\,,
	\label{KeyEq2C}
\end{align}
and 
\begin{align}
	&\sum_{l=n+p-2}^{\infty}\Delta^{l}\sum^{l-n-p+2}_{i=0}c_i\,a_{l-i-n-p+2}\,\big[(l-i-p+2)(l+1)
	+\tfrac{1}{6}(i+p)(i+p-1)\big] \nonumber \\
	& \hspace{50mm} +\tfrac{1}{3}\sum^{\infty}_{l=n}\Delta^{l}\,a_{l-n} = \tfrac{2}{3}\Lambda
	\sum^{\infty}_{l=3n}\Delta^{l}\sum^{l-3n}_{j=0}\sum^{j}_{i=0}a_i\,a_{j-i}\,a_{l-j-3n} \, ,
	\label{KeyEq3C}
\end{align}
respectively. 

Leading orders of equations \eqref{KeyEq1C}-\eqref{KeyEq3C} can be used to determine constraints on the leading powers  $n$ and $p$ in \eqref{rozvojomeg0} and \eqref{rozvojcalH0}.
Since the field equations (\ref{Eq1C})-\eqref{traceC} (and thus also \eqref{KeyEq1C}-\eqref{KeyEq3C}) differ from the vacuum case studied in \cite{PraPraPodSva21} by a single charge-dependent term in (\ref{Eq2C})
the analysis is similar to that done in the vacuum case in \cite{PraPraPodSva21}. Thus, here, we only present Table \ref{tbl1} containing the resulting classes of solutions, while the derivation of these results is given in Appendix \ref{expansiont_0}. Note that the physical interpretation of some of these classes in vacuum has been discussed in \cite{PraPraPodSva21}.

\begin{table}[h]
		\begin{center}
			\begin{tabular}{|c||c|c|c|}
				\hline
				$[n,p]$	& constraints
				& free parameters & physical region  \\[0.5mm]
				\hline\hline
				$[-1,0]$ & $\Lambda\neq0$, $c_0=\frac{\Lambda}{3}a_0^2$ & $a_0$, $c_3$, $r_0$, $q$, $\Lambda$
				& $\bar r \rightarrow \ \infty $ \\
				& $a_1,a_2,a_3,c_1=0$, $c_2=-1$ & &\\
				[1mm]
				$[n<0,$ & $\Lambda=-\frac{3(11n^2+6n+1)}{8k(4n^2-1)}$, & 
				& $\bar r \rightarrow \ \infty $ \\ 
				$p=2n+2<2]$ & $(11n^2+6n+1)c_0=2\Lambda a_0^2$ & &\\[1mm]
				$[-1,2]$	& $\Lambda=0$, $c_0=-1$	& $a_0$, $c_1$, $r_0$, $q$
				&  $\bar r \rightarrow \infty$ \\[1mm]
				$[0,0]$ & &  $a_0$, $a_1$, $c_0$, $c_1$, $c_2$, $r_0$, $q$, $\Lambda$ 
				& $\bar r \rightarrow \bar r_0=a_0 $  \\[1mm]
				$[0,1]$	& 	& $a_0$, $c_0$, $c_1$, $r_0=r_h$, $q$, $\Lambda$
				& $\bar r \rightarrow \bar r_h=a_0 $ \\[1mm]
				$[0,2] $ & $c_0=2\Lambda a_0^2-1$, & & $\bar r \rightarrow \bar r_h=a_0 $\\
				& $2a_0^2(3-8k\Lambda)(1-\Lambda a_0^2)=3\kappa'q^2$ &  $c_1$, $r_0=r_h$, $q$, $\Lambda$ 
				& \\[1mm]
				$[0,p>2]$	& $\Lambda=\frac{3}{2(4k+3\kappa'q^2)}$, $a_0^2=\frac{1}{2\Lambda}$	& 
				& $\bar r \rightarrow \bar r_h=a_0 $\\[1mm]
				$[1,0]$ & & $a_0$, $c_0$, $c_1$, $c_2$, $r_0$, $q$, $\Lambda$ 
				& $\bar r \rightarrow 0 $\\[1mm]
				$[n> 0, 2]$ & $c_0^2 = 1 + \frac{3\kappa'}{4k} q^2$, & 
				& $\bar r \rightarrow 0 $\\
				& $ (3n^2 + 3n + 1)c_0 = -1$ for $\frac{\kappa'}{k} > -\frac{4}{3q^2}$, & 
				& \\
				\hline
			\end{tabular} \\[2mm]
			\caption{Summary of allowed classes $[n,p]$ for power series solutions  \eqref{rozvojomeg0},
				\eqref{rozvojcalH0} to the field equations  (\ref{Eq1C}) and (\ref{Eq2C}). Note that we used only the leading orders of the field equations and thus further analysis may lead to additional restrictions for some of these classes. }
			\label{tbl1}
		\end{center}
	\end{table}


 \section{{(A)dS-Schwarzschild metric}}
\label{sec_Schwarzschild}

The simplest black-hole solution of the vacuum quadratic gravity field equations is the (A)dS-Schwarzschild metric
\eqref{physmet} with
\be
h=f=1-\frac{2M}{\bar r}-\frac{\L}{3}\bar r^2\,,
\ee
which in the Kundt coordinates \eqref{BHmetric} reads
\bea
\H \rovno \frac{\L}{3} -r^2-2M r^3\,,\\
\Omega \rovno -\frac{1}{r}\,.\label{OmSchw}
\eea

For the purposes of this paper, let us express these metric functions as 
expansions around a point $r_0$ (${\Delta=r-r_0}$).\\
$\bullet$ $[0,0]$ solution - expansions around a generic point $r_0$
\bea
\H \rovno \left(\frac{\L}{3} -r_0^2-2M r_0^3\right)
-2r_0 (1+3 M r_0) \Delta-(1+6M r_0)\Delta^2-2 M \Delta^3\,,\label{SchwExpanH}\\
\Omega \rovno -\frac{1}{r_0}
+\frac{\Delta}{r_0^2} -\frac{\Delta^2}{r_0^3}+\frac{\Delta^3}{r_0^4}\dots\,;
\label{SchwExpanOm}
\eea
$\bullet$ $[0,1]$ solution - expansions around a horizon $r_0=r_h$; $\H$ and $\Omega$ are given by \eqref{SchwExpanH}
and \eqref{SchwExpanOm},
where in addition
\be
\frac{\L}{3} -r_h^2-2M r_h^3=0\,;
\label{SchwExpHor}
\ee
$\bullet$ $[0,2]$ solution - expansions around an extremal horizon $\bar r_h
=\frac{1}{\sqrt{\L}}$;  $\H$ and $\Omega$ are given by \eqref{SchwExpanH} and \eqref{SchwExpanOm}, where
\be
r_0=r_h=-\sqrt{\L}\,,\ \ 
M=-\frac{1}{3r_h}\,.
\ee
This is the Kottler metric.

Its near-horizon limit
\bea
\H \rovno \Delta^2 \,,\\
\Omega \rovno \frac{1}{\sqrt{\L}}\,,
\eea
is the Nariai spacetime belonging
to the Kundt family.

\subsection{Number of horizons in the (A)dS-Schwarzschild metric}
\label{SchHor}

Let us study the number of horizons of the (A)dS-Schwarzschild metric for various values of parameters. 
Although for the (A)dS-Schwarzschild metric the  Bach parameter $b$ vanishes, 
we expect qualitative similarities for black holes with sufficiently small $b$.

Let $r=r_0=r_h$ be a horizon. Then the (A)dS-Schwarzschild metric \eqref{BHmetric} in a form convenient for our purposes reads (c.f. \cite{PraPraPodSva21})\footnote{Note that in contrast with \eqref{SchwExpanH}
and \eqref{SchwExpanOm}, here the gauge
$a_0=-1/r_h,\ \  c_0=(1-\L a_0^2) r_h $, leading to \eqref{OmSchw},
is not used.}
\bea
  {\cal H} &=& c_0 \Delta + c_1 \Delta^2 + c_2 \Delta^3 
      = \Delta (c_0 + c_1 \Delta + c_2 \Delta^2)\,, \ \ \ \ \Delta = r-r_h\,,\label{H_Schw[0,1]}\\
      \Omega &=& \frac{a_0 c_0}{(1-\Lambda {a_0}^2) \Delta+ c_0}\,, \label{Omeg_Schw}
\eea
where
\bea
  c_1 &=& 2 - \Lambda a_0^2\,, \\
  c_2 &=& \frac{1}{3c_0}\,  (1-\Lambda a_0^2)(3-\Lambda a_0^2)\,.
\eea
Besides $r_h$, ${\cal H}$ may admit additional real roots.
In such a case, the discriminant of the corresponding quadratic equation has to be non-negative
\begin{equation}
  \Lambda (4-\Lambda a_0^2) \geq 0,
\end{equation}
and thus the additional real roots of $ {\cal H}$  exist iff
\begin{equation}
  0 \leq \Lambda a_0^2 \leq 4.\label{eq2root}
\end{equation}

Recall that the horizon in the Kundt coordinates $r=r_h$ corresponds to $\bar r=a_0$ in the physical coordinates. We will thus further assume $a_0>0$.

Using \eqref{H_Schw[0,1]} and \eqref{Omeg_Schw}, the other roots of $\H $ correspond to  
\begin{equation}
  \bar r_{1,2}   = -\frac{a_0}{2} \mp \frac{\sqrt{3}}{2\,\Lambda a_0}
        \sqrt{\Lambda a_0^2(4-\Lambda a_0^2)}\,.
\end{equation}
The Vieta relations give
\bea
  \bar r_1 + \bar r_2 &=& -a_0<0\,,\\
  \bar r_1 \bar r_2   &=&\frac{\Lambda a_0^2 -3}{\Lambda}\,.
\eea
Thus, using also \eqref{eq2root}, $\bar r_{1}$ is always negative, while $\bar r_{2}$ is positive 
iff 
\begin{equation}
  0<\Lambda a_0^2 < 3. 
\end{equation}
Therefore, besides the horizon at $\bar r=a_0$, there is one additional horizon at $\bar r=\bar r_2$ iff  ${0<\Lambda a_0^2 < 3}$.

   Now, let us study the character of the horizon at ${\bar r=a_0}$. As discussed in \cite{PraPraOrt23},
   this horizon is a black-hole horizon iff 
   \be
   c_0 a_1<0\,.\label{cond_hor}
   \ee
   For the (A)dS-Schwarzschild black hole, 
   \be
 a_1 =-a_0 \frac{1-\Lambda a_0^2}{c_0}\,.
\ee
Thus, $\bar r=a_0$ is a black-hole horizon iff
\be
\Lambda a_0^2<1\,.
\ee

Comparing 
with the standard form of the (A)dS-Schwarzschild metric in the Kundt coordinates \cite{PraPraPodSva21},
we obtain the mass of the (A)dS-Schwarzschild black hole 
\be
m=\frac{a_0}{6}(3-\Lambda a_0^2)\,.
\ee
Thus, $m>0$ for $\Lambda a_0^2<3$. 
For $\Lambda a_0^2>3$, $m<0$, and the solution describes a naked singularity sitting inside a static patch bounded by the cosmological horizon $a_0$.

Let us summarize the dependence of the properties of the  (A)dS-Schwarzschild solution on the value of $\Lambda a_0^2$.

\begin{align}
& & \text{horizon(s)     }& \text{    and their character} & \text{mass} \nonumber\\
\Lambda a_0^2 & <0 & a_0 \ \ \ \ \ \ \  &a_0\text{  BH h., }& m>0\,,\nonumber\\
 0<\,\Lambda a_0^2 & < 1 & a_0 < \bar r_2 \quad \quad 
  & a_0 \text{ BH h., } \bar r_2\ \text{ cosm. h., } & m>0\,,\nonumber\\
  1<\,\Lambda a_0^2 &<3 & a_0 > \bar r_2\quad \quad 
  & a_0 \text{ cosm.\ h., } \bar r_2\ \text{ BH h., } & m>0\,,\nonumber\\
  \Lambda a_0^2 & >3 & a_0 \ \ \ \ \ \ \  &a_0 \text{  cosm. h., } & m<0\,.\nonumber
\end{align}

There are also special cases\\
$\bullet$ $\Lambda a_0^2=0$ is the Schwarzschild solution without a cosmological constant;\\
$\bullet$ $\Lambda a_0^2=1$ is the Kottler spacetime with the extremal horizon. Note that this is a $[0,2]$ solution and thus most of the relations in this section do not apply; \\
$\bullet$ $\Lambda a_0^2=3$  corresponds to the de Sitter spacetime (here $m=0$).

    
	\section{Charged black holes with non-extremal horizons (case [0,1])}
    \label{sec_[0,1]}

In Appendix \ref{expansiont_0}, we derive several possible classes of solutions of static spherically symmetric spacetimes in quadratic gravity in the presence of electromagnetic charge. The necessary conditions for extremality   (the case $[0,2]$)\footnote{Note that $c_0$ in Eq. \eqref{c0[0,2]} is the leading coefficient of $\H$ in the $[0,2]$ case and is distinct from $c_0$ in the $[0,1]$ case, used in the rest of this section.}
derived in Appendix \ref{sec_caseIII} read
\bea
c_0\rovno 2\Lambda a_0^2-1\,,\label{c0[0,2]}\\
2a_0^2(3-8k\Lambda)(1-\Lambda a_0^2)\rovno 3\kappa'q^2\,.\label{qextr}
\eea
This is in agreement with the conditions given in the numerical paper \cite{Lin2016}.

In the rest of this paper, we will focus on the physically most interesting case admitting a non-extremal horizon.

	The lowest nontrivial orders of the trace equation \eqref{KeyEq3C} and  of \eqref{KeyEq2C}  give
	\bea
		a_1 &=& \frac{a_0}{3c_0}\left[2\Lambda a_0^2-(c_1+1)\right] \, ,
		\label{nonSchwinitcond3} \\
	c_2 &=& \frac{1}{6kc_0}\left[2k(c_1^2-1)+a_0^2(2-c_1-\Lambda a_0^2)\right] -\frac{q^2 \kappa'}{4 c_0 k}\, ,
	\label{nonSchwinitcond2}
	\eea
respectively.	
	
	Arbitrary higher orders can be expressed using the recurrent formulas
	\bea
	c_{l+2}\!\!\!&=&\!\!\!\frac{3}{k\,(l+3)(l+2)(l+1)l}\,\sum^{l}_{i=0}a_i \,
	a_{l+1-i}(l+1-i)(l-3i)  \qquad \forall\ l\ge 1\,,
	\label{nonSchwinitcondc}\\
	a_{l}\!\!\!&=&\!\!\!\frac{1}{l^2c_0}\Bigg[\tfrac{2}{3}\Lambda \sum^{l-1}_{j=0}{a_{l-1-j}}\sum^{j}_{i=0}a_i\,a_{j-i}-\tfrac{1}{3}\, a_{l-1}
	-\sum^{l}_{i=1}c_i\,a_{l-i}\left[l(l-i)+\tfrac{1}{6}i(i+1)\right]\Bigg] \ \ \
	\forall \ l\geq 2\,,  \label{nonSchwinitconda}
	\eea
	where $a_0$, $c_0$, $c_1$, and $q$ are arbitrary constants. Note that the form of the above recurrent expressions is identical to the uncharged case \cite{PraPraPodSva21,Svarcetal18}, charge enters implicitly via $c_2$. 
	
	On the horizon, the Bach invariant reads
	\be
	B_{ab}\,B^{ab}(r_h) = \left(\frac{2 a_0^2 ({c_1}-2+ a_0^2 \Lambda)+3 \k 
		q^2}{12 a_0^4 k}\right)^2  =  \left(\frac{2 a_0^2 b +  \k q^2}{4 a_0^4 k}\right)^2 ,
	\label{BachInvariant}
	\ee
	where, similarly as in the vacuum case \cite{PraPraPodSva21,Svarcetal18}, we introduce a {dimensionless} Bach parameter $b$ by
	\be
	b \equiv \frac{1}{3}\left(c_1-2 +\Lambda a_0^2\right)\,, \label{b_definice}
	\ee
	corresponding to the strength of the Bach tensor at the horizon when $q=0$.

    Using \eqref{b_definice}, the expressions \eqref{nonSchwinitcond3} and \eqref{nonSchwinitcond2}
    go to
    \bea
    a_1\rovno \frac{a_0}{c_0}(\L a_0^2 -1-b)\,,\label{nonSchwinitcond3b}\\
    c_2\rovno  \frac{1}{6kc_0}\left[2k  (3 b+3 -\Lambda a_0^2 )(3 b+1 -\Lambda a_0^2)
    -3ba_0^2\right] -\frac{q^2 \kappa'}{4 c_0 k}\,.\label{nonSchwinitcond2b}
	\eea


The Weyl  invariant on the horizon reads
\be
{C_{abcd}}C^{abcd}\rightarrow\frac{4(c_1+1)^2}{3a_0^4}\,.
\ee

It has been shown \cite{PraPraTur24}  that in the case of a vanishing cosmological constant, quadratic gravity admits two static spherically symmetric, asymptotically flat charged black holes:
the charged Schwarzschild
	 and charged Schwarzschild-Bach solutions. 
To identify which of the two above classes a charged black hole belongs to, one has to determine whether, in the $q=0$ limit, the black hole approaches Schwarzschild or Schwarzschild-Bach. This is non-trivial and involves fine-tuning for asymptotic flatness for gradually decreasing values of $q$ (see \cite{PraPraTur24}). As will be seen in Sec. \ref{tuning_elmg}, in the presence of a cosmological constant, the distinction between the charged Schwarzschild and charged Schwarzschild-Bach solutions becomes less clear.	

It turns out that for sufficiently large values of $|\Lambda|$, the asymptotic behaviour of static spherically symmetric black holes in quadratic gravity is qualitatively different from the $\L=0$ case. There exists an interval of values of the Bach parameter $b$ such that they are asymptotically (A)dS,  without the need for fine-tuning. Thus, within this interval, the Bach parameter becomes one additional free physical parameter (see also numerical results of \cite{Luetal12} for conformal and Einstein-Weyl gravity in vacuum and a brief discussion in \cite{PraPraOrt23}).  

Let us, in this context, study this behaviour for both charged and uncharged black holes.

\begin{figure}[h!]
	\centering
	\includegraphics[height=54mm]{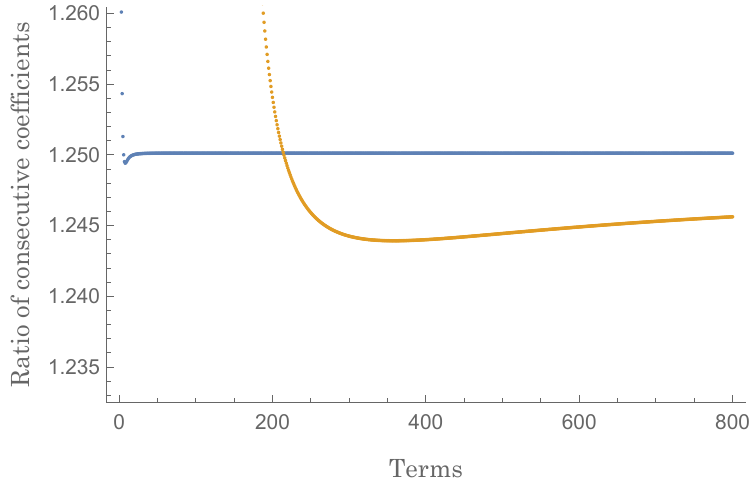}
	\includegraphics[height=54mm]{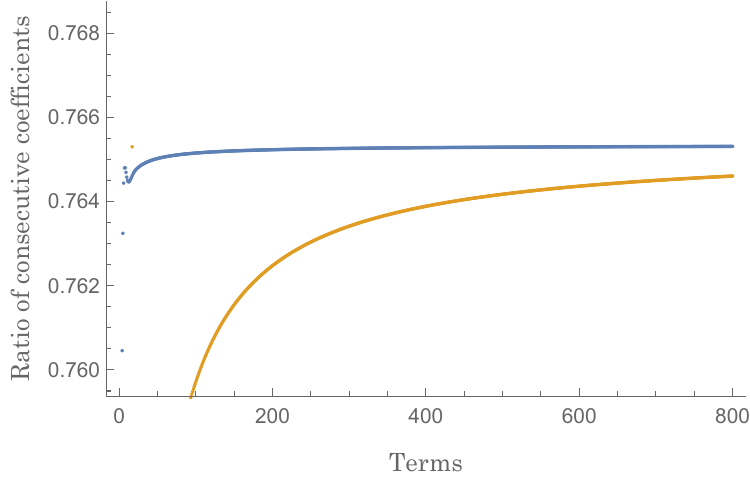}
	\caption{Ratios of consecutive coefficients, $a_i/a_{i-1}$ (blue) and $c_i/c_{i-1}$ (yellow), for $[0,1]$ solutions with $r_h=-1$, $k=1/2$, and $\kappa'=1$. The remaining parameters are $q=0.3$, $b=-0.2$, $\Lambda=2$, $a_0=1$, $c_0=1$ (left) and  $q=0.8$, $b=-0.4$, $\Lambda=-1$, $a_0=1$, $c_0=-2$ (right).}
	\label{fig:terms}
\end{figure}

\begin{figure}[h!]
	\centering
	\includegraphics[height=54mm]{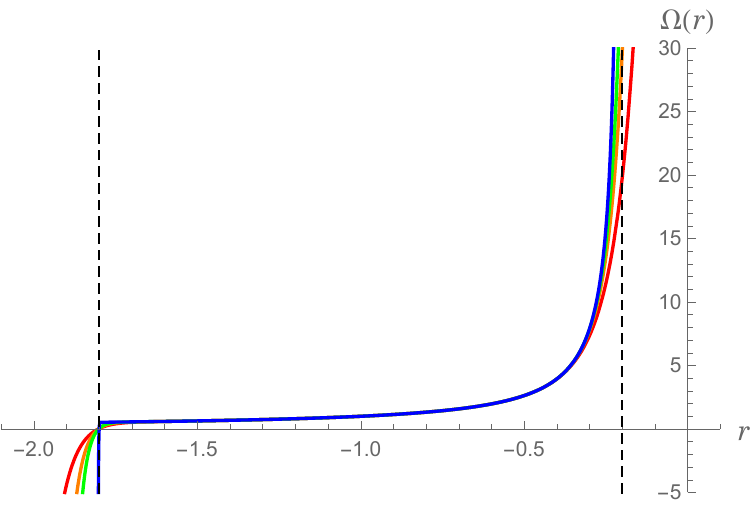}
	\includegraphics[height=54mm]{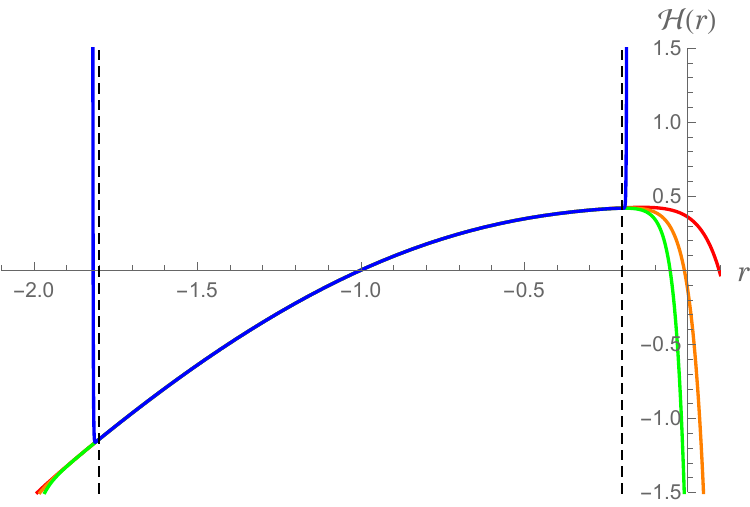}
	\caption{The metric functions $\Omega(r)$ and $\mathcal{H}(r)$ for the $[0,1]$ solution with $r_h=-1$, $k=1/2$, $\kappa'=1$, $q=0.3$, $b=-0.2$, $\Lambda=2$, $a_0=1$, and $c_0=1$. Expansions with 20 (red), 30 (orange), 40 (green), and 800 (blue) terms are shown with the black dashed lines indicating the interval of convergence.}
	\label{fig:omega}
\end{figure}

\begin{figure}[h!]
	\centering
	\includegraphics[height=54mm]{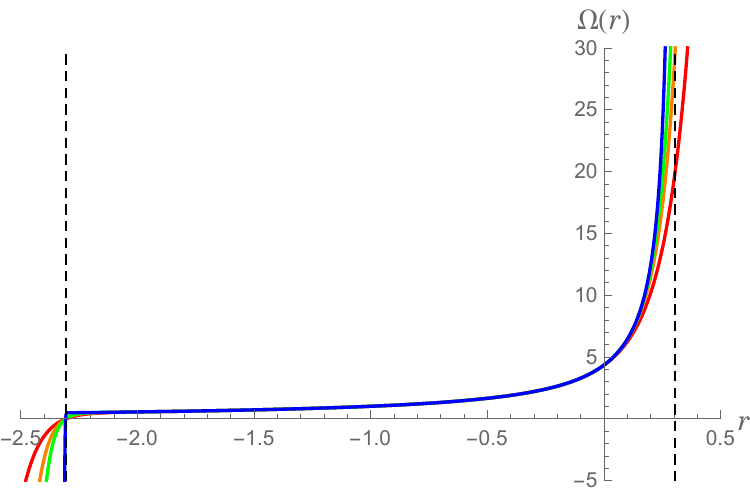}
	\includegraphics[height=54mm]{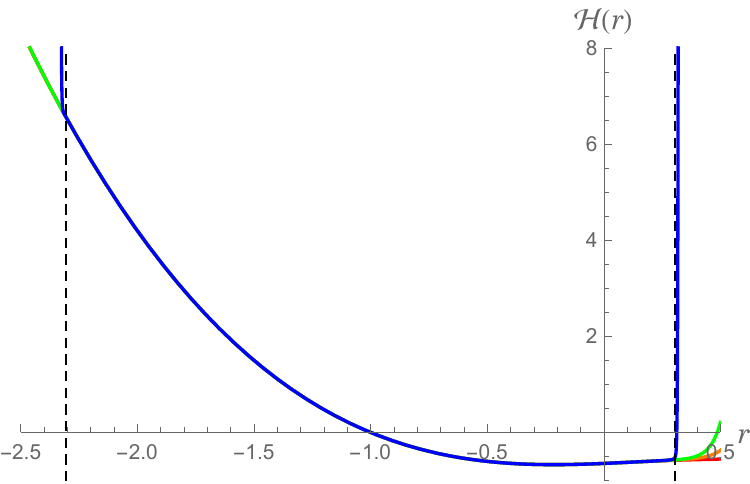}
	\caption{The metric functions $\Omega(r)$ and $\mathcal{H}(r)$ for the $[0,1]$ solution with $r_h=-1$, $k=1/2$, $\kappa'=1$, $q=0.8$, $b=-0.4$, $\Lambda=-1$, $a_0=1$, $c_0=-2$. Expansions with 20 (red), 30 (orange), 40 (green), and 800 (blue) terms are shown with the black dashed lines indicating the interval of convergence.}
	\label{fig:omega2}
\end{figure}

\begin{figure}[tbh!]
	\centering
	\includegraphics[height=52mm]{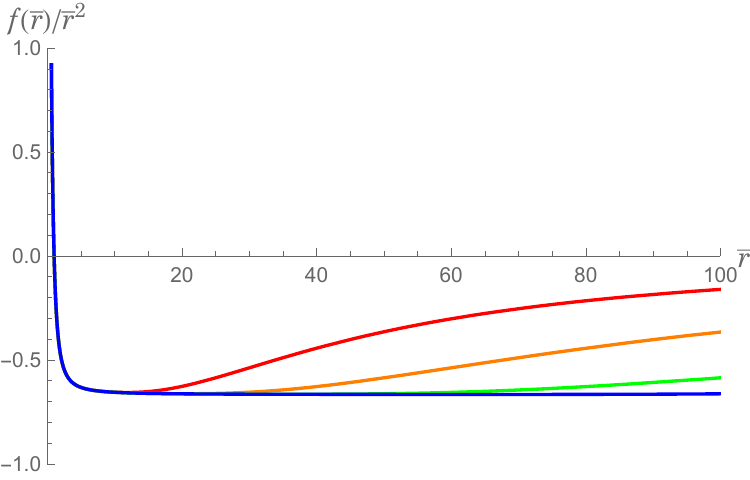}
	\includegraphics[height=52mm]{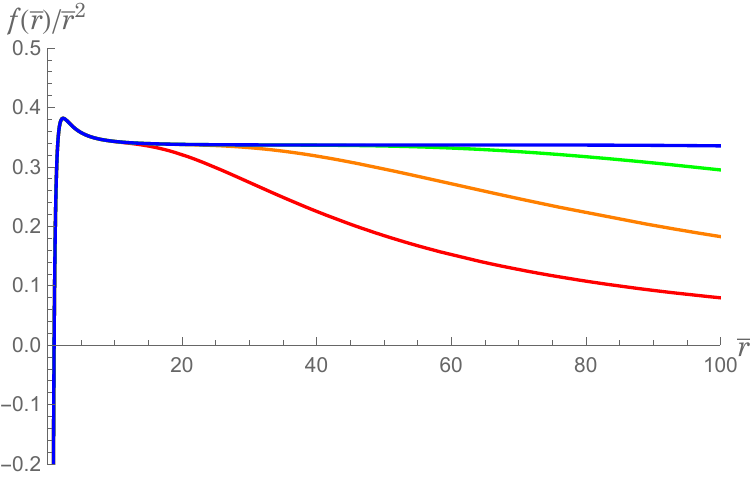}
	\caption{Plot of $f(\bar r)/\bar r^2$ illustrating asymptotic behaviour of $[0,1]$ solutions with $r_h=-1$, $k=1/2$, and $\kappa'=1$. Expansions with 100 (red), 200 (orange), 300 (green), and 800 (blue) terms are shown for $q=0.3$, $b=-0.2$, $\Lambda=2$, $a_0=1$, $c_0=1$ (left) and $q=0.8$, $b=-0.4$, $\Lambda=-1$, $a_0=1$, $c_0=-2$ (right).}
	\label{fig:frr}
\end{figure}

First, note that in many cases, as in the uncharged case, the series of coefficients $a_i$ and $c_i$ asymptotically approach a geometric series (see Fig. \ref{fig:terms}).
This is useful for an estimate of the convergence interval for the metric functions $\Omega$ and ${\cal H}$ indicated by vertical dashed lines in Figs. \ref{fig:omega} and \ref{fig:omega2} for a positive and negative value of $\L$, respectively.
The corresponding metric function $f$ as a function of the physical coordinate $\bar r$ can be obtained from $\Omega$ and ${\cal H}$ via \eqref{to static} and \eqref{Schwarz}, see Fig. \ref{fig:frr}.
\clearpage

\section{Exploring the asymptotically (A)dS regions of the parameter space in the vacuum case}
\label{tuning_vac}

Let us now clarify which regions of the parameter space are asymptotically (A)dS with/without fine-tuning. In these regions, either fine-tuning of $b$ is needed to achieve asymptotic (A)dS-ness or there is a continuous interval of values of the Bach parameter $b$ compatible with (A)dS asymptotics, respectively.

\begin{figure}[h!]
	\centering
	\includegraphics[height=70mm]{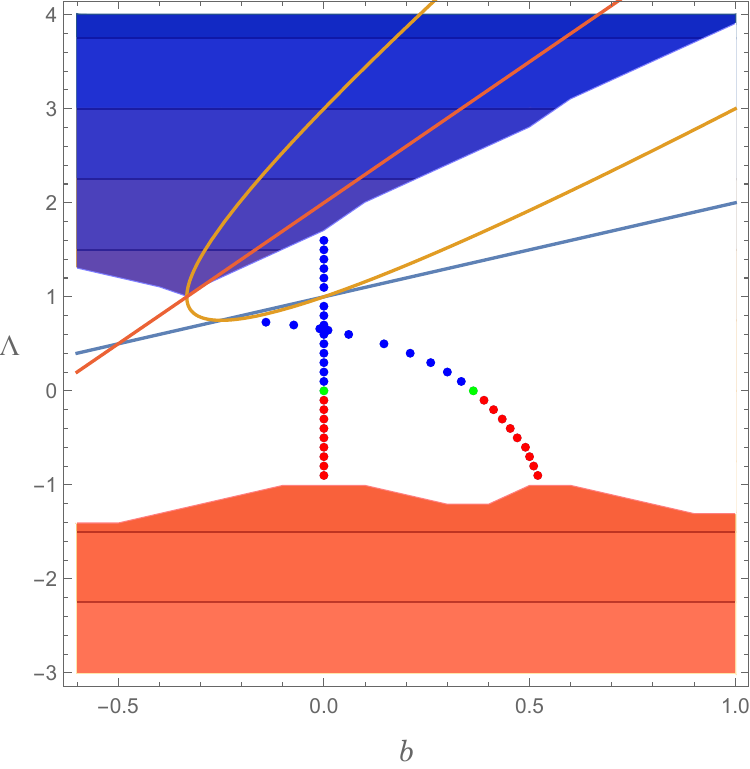}
	\caption{Plot of $\Lambda$ vs $b$ for $[0,1]$ solutions with common parameters $r_h=-1$, $k=1/2$, $\kappa'=1$, $q=0$, $a_0=1$, and $c_0=-1$. The blue and red regions contain asymptotically de Sitter and anti-de Sitter solutions respectively, with $f(\bar r)/\bar r^2$ approaching $-\Lambda/3$, 
    in each case without the need for fine-tuning.
    These are G(A)dS Schwarzschild--Bach black holes (or for large positive $\L$, these could be naked singularities surrounded by a cosmological horizon, c.f. Sec. \ref{SchHor}). Between these regions, there are two families of fine-tuned solutions with AdS (red), dS (blue), and flat (green) asymptotic behaviour. 
    In addition, the plot contains several curves indicating 
     where various early terms in the expansions vanish:  $a_1=0$ on the blue line ($\Lambda=b+1$); $c_1=0$ on the orange line ($\Lambda=3b+2$); $c_2=0$ on the yellow line ($\Lambda=3b+2\pm\sqrt{3b+1}$). 
     Note that below the blue line, solutions are expanded around the black-hole horizon, while above it, they are expanded around the cosmological horizon.
     Intersections of these lines are discussed in the main text. }
	\label{fig:q0Lambda}
\end{figure}

First, let us look at Fig. \ref{fig:q0Lambda} describing these regions in the uncharged case, which was studied in \cite{Svarcetal18,PraPraPodSva21}, however, asymptotic properties were not discussed there. 

In the red and blue regions, the solutions are asymptotically AdS and dS, respectively, and fine-tuning is not needed.
In contrast, in the white region, fine-tuning for asymptotic flatness or asymptotic (A)dS-ness is necessary, and one can identify two branches of asymptotically (A)dS solutions there
for given values of $a_0$ and parameters of the theory. The vertical $b=0$ dotted line is the (A)dS-Schwarzschild black hole.  The second dotted curve in the white region is the FT(A)dS–Schwarzschild–Bach black hole, for which the Bach parameter $b$ has been fine-tuned to achieve asymptotic (A)dS-ness for each value of $\Lambda$ separately.

\begin{figure}[h!]
    \centering
    \includegraphics[height=52mm]{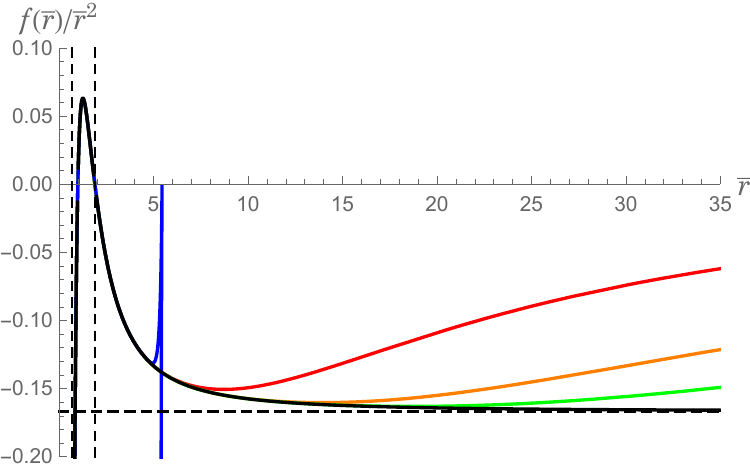}
    \includegraphics[height=52mm]{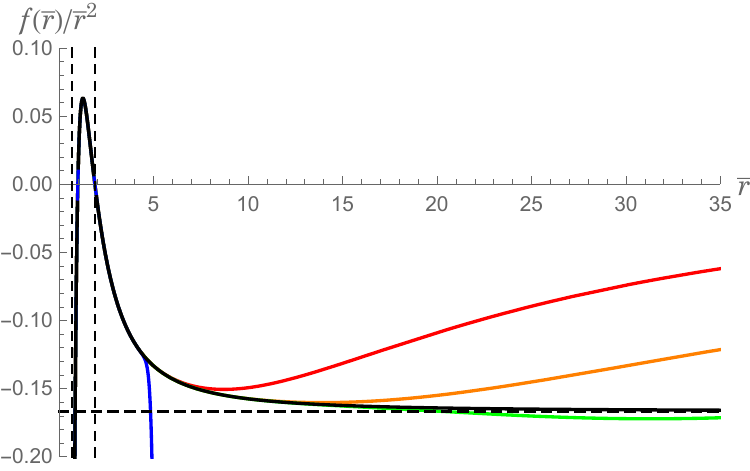}
    \caption{Plot of $f/\bar r^2$ for tuned $[0,1]$ solutions illustrating the tuning process and comparison with the numerical solution. Common parameters are $r_h=-1$, $k=1/2$, $\kappa'=1$, $q=0$, $\Lambda=1/2$, $a_0=1$, and $c_0=-1$. On the left $b=1460455736505795379977859245805530483862308/\
10^{43}$ and on the right $b=1460455736505795379977859245805530483862307/\
10^{43}$. In each plot, the number of terms in the expansion is varied between 50 (red), 100 (orange), 150 (green), and 200 (blue). The solid black lines indicate patches of numerical solutions starting on either side of both horizons
(glued together across horizons using the series solution). Obviously, the infinite series diverges abruptly outside the interval of convergence (the black, dashed, vertical line that approaches the second horizon as the number of terms increases). An appropriate finite number of terms, however, can be fine-tuned to approach the asymptotic value of $-\Lambda/3$, indicated by the black, dashed, horizontal line. At each degree of precision of $b$ during the tuning process, there is an optimal truncation of the series (approximated here by the green line at 150 terms). Below this, the red and orange lines do not respond to further tuning, while the blue line at 200 terms is limited by the precision of $b$. Further tuning increases the number of terms required for optimal truncation.}
    \label{fig:dstuning3}
\end{figure}

\begin{figure}[h!]
    \centering
    \includegraphics[height=52mm]{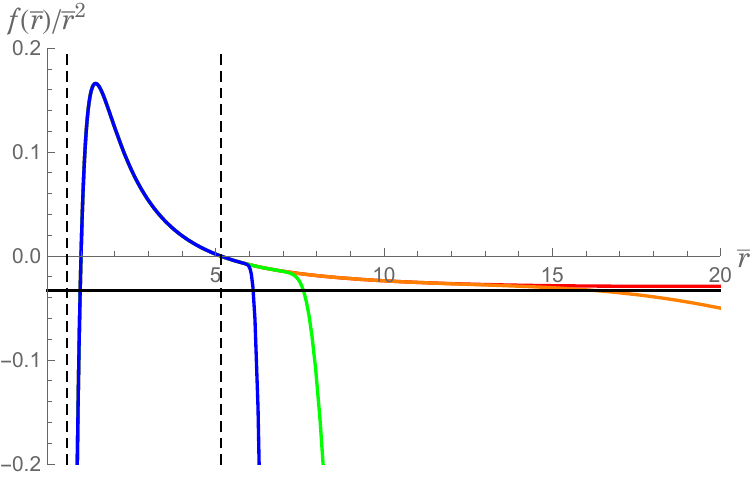}
    \includegraphics[height=52mm]{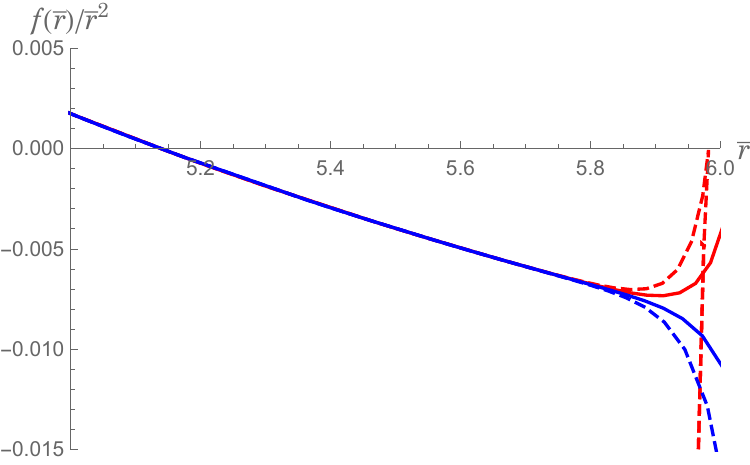}
    \caption{Plot of $f/\bar r^2$ for tuned $[0,1]$ solutions illustrating the tuning process. Common parameters are $r_h=-1$, $k=1/2$, $\kappa'=1$, $q=0$, $\Lambda=0.1$, $a_0=1$, and $c_0=-1$. On the left, $b=0.33400590906606$ is constant, and the number of terms in the expansion varies between 100 (red), 200 (orange), 400 (green), and 800 (blue). Obviously, the infinite series diverges abruptly outside the interval of convergence (the black, dashed, vertical line that approaches the second horizon as the number of terms increases). An appropriate finite number of terms, however, can be fine-tuned to approach the asymptotic value of $-\Lambda/3$, indicated by the solid, black line.  This process is depicted on the right, where the solid blue line is the same solution with $b=0.33400590906606$ and 800 terms. The other lines represent solutions with the same number of terms but with $b$ varied by increments in the last digit. Note that a degree of fine-tuning is required to even reach the second horizon.}
    \label{fig:dstuning2}
\end{figure}

While for $\L<0$, the fine-tuning process remains much the same as in the $\L=0$ case (albeit requiring less precision to reach comparable radii), in other cases, fine-tuning is considerably more involved. 
As discussed in Sec. \ref{sec_specialuncharged}, for solutions with $\L>0$ in the white region above the line $a_1=0$, the expansion is done around the outer (cosmological) horizon. Here, the fine-tuning is qualitatively similar to the $\L=0$ case. In contrast, for solutions with $\L>0$ in the white region below the line $a_1=0$, the black-hole metric is expanded around the black-hole horizon, which is inside the cosmological horizon (see Sec. \ref{sec_Schwarzschild} where this is discussed for (A)dS-Schwarzschild). At a very early stage in the tuning process, the series solution itself may not reach the cosmological horizon at all. Once $b$ is precise enough, the cosmological horizon is always present, and
the radius of convergence of the series approaches the cosmological horizon.   Further tuning is necessary to reach the asymptotic dS property. This is complicated by the fact that the convergence interval of our series still ends on the cosmological horizon. Nevertheless, appropriately truncated series match numerical solutions outside the cosmological horizon very well\footnote{See, e.g., \cite{Boyd99} for a discussion of using truncated formal series solutions to ODEs outside of the interval of convergence.} and can be used to further fine-tune the Bach parameter. This type of fine-tuning is illustrated in Figs. \ref{fig:dstuning3}
and \ref{fig:dstuning2}, where the formal solution is expanded around the black-hole horizon, but the truncated series approximates the numerical solution well beyond the cosmological horizon.

\subsection{Special subcases}
\label{sec_specialuncharged}
Let us now identify special cases appearing in Fig. \ref{fig:q0Lambda}.

{\bf{A) Line $a_1=0$}}

First, let us examine the line $a_1=0$. Using Eq. \eqref{nonSchwinitcond3b}, this implies
\be
\L a_0^2=b+1.\label{a1nula}
\ee
This line divides regions where the $[0,1]$ solution is expanded around the black-hole horizon (below this line) and around the cosmological horizon (above the line), c.f. \eqref{cond_hor}. In point B below, some special points on this line are further discussed.

Eqs. \eqref{nonSchwinitcond3}, \eqref{nonSchwinitcond2}, and \eqref{nonSchwinitconda} imply 
\bea
c_1\rovno 2\L a_0^2-1\,,\label{eq_c1}\\
c_2\rovno \frac{a_0^2(\L a_0^2-1)(8k\L -3)}{6kc_0}\,,\label{eq_c2q0}\\
a_2\rovno -\frac{a_0^3(\L a_0^2-1)(8k\L -3)}{24kc_0^2}\,.\label{eq_a2q0}
\eea

The Bach and Weyl invariants at the horizon read
\bea
	B_{ab}\,B^{ab}(r_h) \rovno \left(\frac{2 a_0^2 (\Lambda a_0^2-1)}{4k a_0^4 }\right)^2\,, 
	\label{BachInvarianta1}\\
	{C_{abcd}}C^{abcd}({r_h})\rovno \frac{16\L^2 }{3}\,.
\eea
Since $a_1$ vanishes, Eq. \eqref{to static} implies that integer steps in $\Delta$ in the power-series solution
translate to steps in $\bar\Delta^{1/2}$ in the physical coordinates. This solution was found in \cite{Luetal15b}, generalized to $\L\not=0$ in \cite{PraPraPodSva21}, and interpreted in \cite{PerkinsPhD} as an unusual horizon.

Note that for $a_1=0$, the conditions $c_2=0$ \eqref{eq_c2q0} and
$a_2=0$ \eqref{eq_a2q0} are equivalent. This special case is discussed in point B.

{\bf{B) Intersections of the curve $c_2=0$
and line $a_1=0$}}

Let us now concentrate on the intersections of the line $a_1=0$ (the blue line in Fig. \ref{fig:q0Lambda} defined by Eq. \eqref{a1nula})
and
the curve $c_2=0$ (the yellow curve in Fig. \ref{fig:q0Lambda}) 
defined by  (see Eq. \eqref{nonSchwinitcond2b})  
\be
(1-\L a_0^2)(3-\L a_0^2)+3\left[4-\left(2\L+\frac{1}{2k}\right)a_0^2\right] b+9b^2=0\,.\label{curvec20}
\ee
The intersection gives
\be
4ba_0^2\left(\L-\frac{3}{8k}\right)=0
\ee
and thus either 
B1) $b=0$, i.e., $\L a_0^2=1$ from \eqref{a1nula},  or 
B2) $\L=\frac{3}{8k}$.

In both cases,  one can show using mathematical induction and the recurrent relations \eqref{nonSchwinitcondc}, \eqref{nonSchwinitconda} that all coefficients $a_i$, $i>0$, and $c_j$, $j>1$, vanish. Consequently, the metric reduces to a closed form
\bea
\H (r)&=&(r -r_h)[c_0+c_1(r -r_h) ]\,,
\label{intersectionbluegreen}\\
\Omega&=&a_0\,,
\eea
which is a Kundt metric.\\
$\bullet$ B1) Case $b=0$ (and from \eqref{a1nula} $\L a_0^2=1$) with $c_1=1$  is the Nariai spacetime  (see Sec. C1 in \cite{PraPraPodSva21}), 
for $c_0=0$, describing the near-horizon limit of extremal dS-Schwarzschild black hole (see Sec. E1 in \cite{PraPraPodSva21} for extremal Schwarzschild in Kundt coordinates) and extremal higher-order (discrete) dS-Schwarzschild-Bach black holes, see Sec. E1 in \cite{PraPraPodSva21}.\\[2mm]
$\bullet$ B2) Case $\L=\frac{3}{8k}$:   $c_1=(2\L a_0^2-1)=2b+1$. For $c_0=0$,  this Kundt metric is a Bachian
generalization of the Nariai spacetime  and corresponds to the near-horizon limit of
another extremal
dS-Bachian black hole, both discussed in Sec. E2
of \cite{PraPraPodSva21}.\\[3mm]

{\bf{C) Intersection of the line $c_1=0$ and the curve $c_2=0$}} 

There is one point in the intersection of the yellow curve $c_2=0$ \eqref{curvec20} and the orange line $c_1=0$, c.f., \eqref{b_definice},
\be
	3 b +2 -\Lambda a_0^2 =0\,,\label{curvec10}
	\ee
given by
\bea
b\rovno -\frac{2k}{3a_0^2}\,,\\
\Lambda a_0^2\rovno 2 -\frac{2k}{a_0^2}\,.
\eea
The metric with these parameters is asymptotically dS.

\FloatBarrier

\section{Exploring the asymptotically (A)dS regions in the charged case}

\label{tuning_elmg}

\begin{figure}[h!]
	\centering
	\includegraphics[height=70mm]{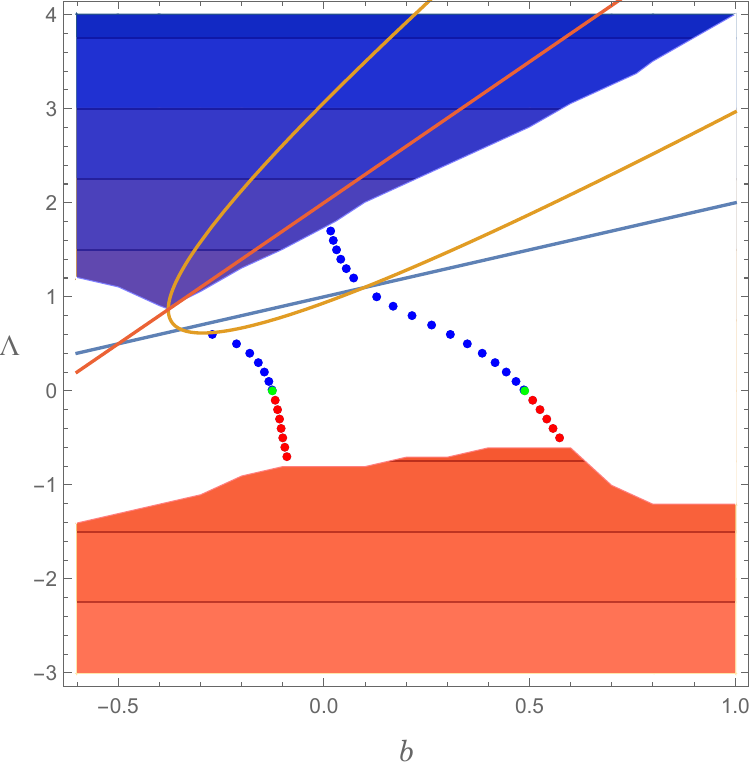}
	\caption{Plot of $\Lambda$ vs $b$ for $[0,1]$ solutions with common parameters $r_h=-1$, $k=1/2$, $\kappa'=1$, $q=0.3$, $a_0=1$, and $c_0=-1$. The blue and red regions contain asymptotically de Sitter and anti-de Sitter solutions respectively, with $f(\bar r)/\bar r^2$ approaching $-\Lambda/3$, in each case without the need for fine-tuning. {These are charged G(A)dS-QG black holes (or for large positive $\L$, these could be naked singularities surrounded by a cosmological horizon, c.f. Sec. \ref{SchHor})}. Between these regions are two families of fine-tuned solutions with AdS (red), dS (blue), and flat (green) asymptotic behaviour. In addition, the plot contains several curves indicating where various early terms in the expansions vanish:  $a_1=0$ on the blue line ($\Lambda=b+1$); $c_1=0$ on the orange line ($\Lambda=3b+2$); $c_2=0$ on the yellow line ($\Lambda=3b+2\pm\sqrt{3b+1+\frac{3}{2}q^2}$). Note that below the blue line, solutions are expanded around the black-hole horizon, while above it, they are expanded around the cosmological horizon. Intersections of these lines are discussed in the main text. }
	\label{fig:q03Lambda}
\end{figure}

Let us now explore asymptotics of charged static spherically symmetric black holes in quadratic gravity, which is illustrated in Fig. \ref{fig:q03Lambda}.
As in the vacuum case (see Sec. \ref{tuning_vac}), fine-tuning for asymptotic  (A)dS-ness is not needed in the red and blue regions (where the solutions are automatically asymptotically AdS and dS, respectively) and is necessary in the white region.

\begin{figure}[h!]
    \centering
    \includegraphics[height=50mm]{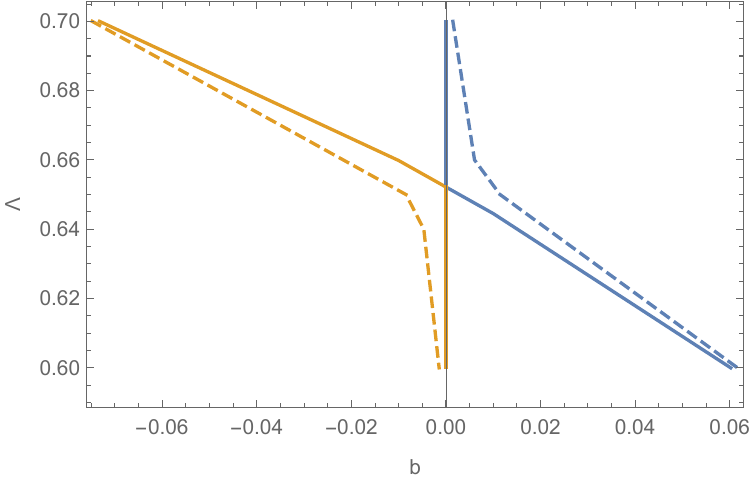}
    \caption{Plot of $\Lambda$ vs $b$ for $[0,1]$ solutions with common parameters $r_h=-1$, $k=1/2$, $\kappa'=1$, $a_0=1$, and $c_0=-1$ (i.e., the vicinity of the intersection of tuned solutions in Fig. \ref{fig:q0Lambda}). The solid lines indicate solutions with $q=0$: dS-Schwarzschild, where $b=0$; and additional fine-tuned solutions with de Sitter asymptotic behaviour that intersect at $\Lambda=\L_{\times}\approx0.65$. The dashed lines indicate the separation of these two families of solutions with the addition of a small charge $q=0.01$. }
    \label{fig:lowqsplit}
\end{figure}

By adding a small charge, the picture qualitatively changes. In the region where fine-tuning is necessary, there are still two branches of asymptotically (A)dS black holes as in the vacuum case. However, they do not intersect, see Figs. \ref{fig:lowqsplit}, \ref{fig:q03Lambda}. 
{In our previous paper on the 
$\Lambda=0$  case \cite{PraPraTur24}, we referred to the two black-hole branches as ``charged Schwarzschild'' and ``charged Schwarzschild–Bach''.}
In the presence of the cosmological constant, the situation is more fuzzy as one branch is close to Schwarzschild for $\L<\L_{\times}$ and close to Schwarzschild-Bach for  $\L>\L_{\times}$ and vice versa for the second branch, see Fig. \ref{fig:lowqsplit}. Thus, we will not use this terminology here.

Similarly to the vacuum case, the fine-tuning of solutions with $\L>0$ in the white region below the line $a_1=0$ is more involved. This is again because the interval of convergence of the series expansion around the black-hole horizon ends on the outer cosmological horizon, and we need the truncated series to complete the fine-tuning.

\begin{figure}[h!]
    \centering
    \includegraphics[height=52mm]{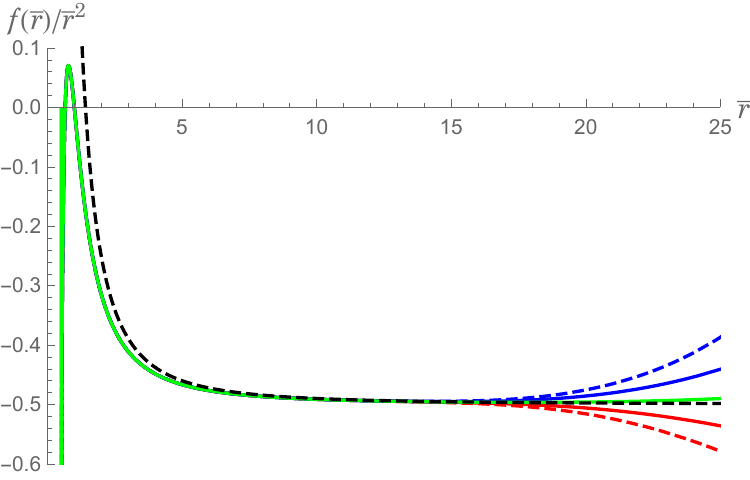}
    \includegraphics[height=52mm]{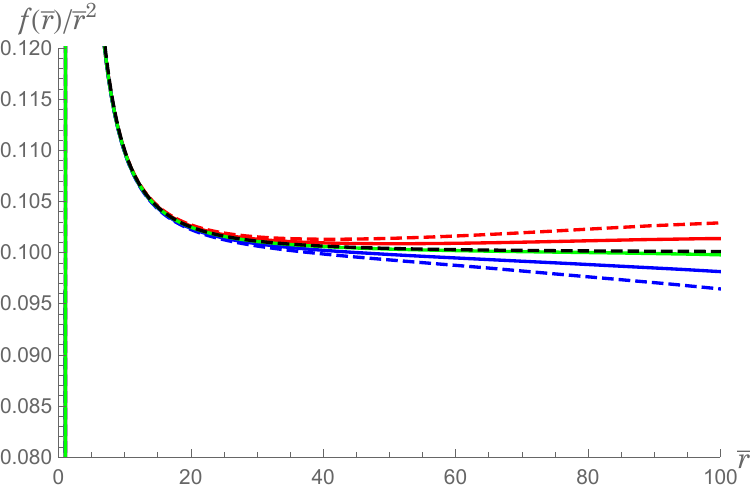}
    \caption{Left: Plot of $f/\bar r^2$ for highly tuned $[0,1]$ solutions illustrating the tuning process. Common parameters are $r_h=-1$, $k=1/2$, $\kappa'=1$, $q=3/10$, $\Lambda=3/2$, $a_0=1$, and $c_0=-1$ with 200 terms in the expansion. The most precise solution is in green with $b=3073799795990587885810658705398202090271506172332/10^{50}$, while the others vary by increments in the last digit. The black dashed line represents $1/r^2-\Lambda/3$. 
    Note that for the solution expanded around the cosmological horizon and fine-tuned for asymptotic dS-ness, the lower bound of the interval of convergence approaches the inner black-hole horizon. Truncated series and numerics can be used to extend the solution below the inner black-hole horizon, similarly to some previous cases, see, e.g., Fig. \ref{fig:dstuning3}.
    Right: Plot of $f/\bar r^2$ for tuned $[0,1]$ solutions illustrating the tuning process. Common parameters are $r_h=-1$, $k=1/2$, $\kappa'=1$, $q=0.3$, $\Lambda=-0.3$, $a_0=1$, and $c_0=-1$ with 1000 terms in the expansion. The most precise solution is in green with $b=0.5418$, while the others vary by increments in the last digit. The black dashed line represents $1/r^2-\Lambda/3$. It should be noted how much easier the tuning process is for lower values of $\Lambda$. }
    \label{fig:dstuning}
\end{figure}

In the white region above the line $a_1=0$, the solution is expanded around the cosmological horizon. Once tuned to asymptotic dS-ness (see Fig. \ref{fig:dstuning}), the lower boundary of the interval of convergence approaches the inner black-hole horizon. Truncated series and numerics can be used to study the solution below the black-hole horizon. Note that the fine-tuning requires much higher precision for higher values of $\L$ in the white region (Fig. \ref{fig:dstuning}).

To complete our discussion of [0,1] static spherically symmetric solutions in electrovacuum quadratic gravity depicted in Fig. \ref{fig:q03Lambda}, in the next section, let us extend the analysis to several special cases.

\subsection{Special subcases}
\label{sec_specialq}
Let us now identify special cases appearing in the Fig. \ref{fig:q03Lambda}.

{\bf{A) Line $a_1=0$}}

First, let us examine the line $a_1=0$. As in the vacuum case, this line divides regions where the $[0,1]$ solution is expanded around the black-hole (below this line) and cosmological horizon (above the line), c.f. \eqref{cond_hor}. Some special points on this line are discussed in point B below.

Relations \eqref{nonSchwinitcond3}, \eqref{nonSchwinitcond2}, and \eqref{nonSchwinitconda} imply 
\bea
c_1\rovno 2\L a_0^2-1\,,\label{eq_c1q}\\
c_2\rovno \frac{2a_0^2(\L a_0^2-1)(8k\L -3)-3\k q^2}{12kc_0}\,,
\label{eq_c2q}\\
a_2\rovno - \frac{a_0[2a_0^2(\L a_0^2-1)(8k\L -3)-3\k q^2]}{48kc_0^2}\,.
\label{eq_a2q}
\eea

The Bach and Weyl invariants at the horizon read
\bea
	B_{ab}\,B^{ab}(r_h) \rovno \left(\frac{2 a_0^2 (\Lambda a_0^2-1)+ \k 
		q^2}{4k a_0^4 }\right)^2\,, 
	\label{BachInvarianta1q}\\
	{C_{abcd}}C^{abcd}({r_h})\rovno \frac{16\L^2 }{3}\,.
\eea

Since $a_1$ vanishes, Eq. \eqref{to static} again implies that 
integer steps in $\Delta$
in the power series solution
translate to  steps in $\bar\Delta^{1/2}$ in the physical coordinates.  
 This solution is the charged generalization of the ``unusual horizon'' solution discussed in Sec. \ref{sec_specialuncharged}, case A.

Note that for $a_1=0$, the conditions  $c_2=0$ \eqref{eq_c2q}
and $a_2=0$ \eqref{eq_a2q} are equivalent. This special case is discussed in point B.


{\bf{B) Intersection of the curve $c_2=0$
and line $a_1=0$}}

The intersection of  the blue line $a_1=0$ (see \eqref{a1nula}) in Fig. \ref{fig:q03Lambda} and the yellow curve $c_2=0$ therein (see eq. \eqref{nonSchwinitcond2b})
 \be
     (3 b+3 -\Lambda a_0^2 )(3 b+1 -\Lambda a_0^2)
    -\frac{3ba_0^2}{2k} -\frac{3 q^2 \kappa'}{4  k}=0\,.\label{curvec20q}
	\ee
The intersection thus gives two solutions (note that this is the extremal condition \eqref{qextr})
\be
4ba_0^2\left(\L-\frac{3}{8k}\right) -\frac{3 q^2 \kappa'}{4  k}=0 \label{01specialintersa1c2}
\ee
or using \eqref{a1nula},
\be
4b\left(b+1-\frac{3a_0^2}{8k}\right) -\frac{3 q^2 \kappa'}{4  k}=0 \,.\label{01qeqb}
\ee
Then
\be
b_{1,2}=\frac{-(1-\frac{3a_0^2}{8k})\pm\sqrt{(1-\frac{3a_0^2}{8k})^2+\frac{3 q^2 \kappa'}{4  k}}}{2}\,.\label{sol_b}
\ee
Note that in the $q=0$ limit, the $b=b_1$ case is the near-horizon limit of the extremal Schwarzschild, see case B1 in Sec. \ref{sec_specialuncharged}, 
while the $b=b_2$ reduces to the B2 case of Sec. \ref{sec_specialuncharged}.

As in the uncharged case, one can show that
 all the remaining coefficients $a_i$, $i>0$, and $c_j$, $j>1$, vanish. Thus, the metric reduces to
\bea
\H (r)&=&(r -r_h)[c_0+c_1(r -r_h) ]\,,
\label{intersectionbluegreen}\\
\Omega&=&a_0\,,
\eea
where $c_1=2b+1$, with 
$b$ given by \eqref{sol_b}.
For $c_0=0$, these solutions are the near-horizon limits of extremal charged black holes (charged generalizations of the cases B1 and B2 in Sec. \ref{sec_specialuncharged}). Both solutions belong to the Kundt family.

{\bf{C) Intersection of the line $c_1=0$ and the curve $c_2=0$}}

The point at the intersection of the yellow curve $c_2=0$ \eqref{curvec20q} and the orange line $c_1=0$
\eqref{curvec10}
\bea
b\rovno -\frac{2k}{3a_0^2}
- \frac{  q^2 \kappa'}{2  a_0^2}\,,\\
\Lambda a_0^2\rovno 2 -\frac{2k}{a_0^2}
- \frac{  3 q^2 \kappa'}{2  a_0^2}\,
\eea
corresponds to a  $[0,1]$ solution with dS asymptotics.

\section*{Acknowledgement}
This work has been supported by the Institute of Mathematics, Czech Academy of Sciences (RVO 67985840) and research grant GA25-15544S.

   \appendix

\section{ Constraints on $n$ and $p$ from indicial equations}
\label{expansiont_0}

In this section, we will look at the leading orders of equations 
\eqref{KeyEq1C}-\eqref{KeyEq3C} and determine constraints on the leading powers  $n$ and $p$ in  (\ref{Eq1C}) and (\ref{Eq2C}). Note that some of the classes we obtain in this section may be empty, since while being compatible with leading terms, they may be incompatible with higher-order terms in the expansion of the field equations.

 Since the lowest
orders on the left and right hand sides of \eqref{KeyEq1C} are ${l=2n-2}$ and ${l=p-4}$,
respectively, it is convenient to split the solutions into three classes
\begin{itemize}
	\item Case I: ${\ \ 2n-2<p-4}$\,, \ i.e.,  ${\ p>2n+2}$\,,
	\item Case II: ${\ 2n-2>p-4}$\,,  \ i.e.,  ${\ p<2n+2}$\,,
	\item Case III: ${2n-2=p-4}$\,,   \ i.e.,  ${\ p=2n+2}$\,.
\end{itemize}

 Given that Eq.~\eqref{KeyEq1C} does not contain $\L$ nor $q$, the classes are same as in the case with no $\L$ and $q$
  \cite{PodSvaPraPra20}. Since the charge $q$ appears only in  \eqref{KeyEq2C}, the subsequent analysis of the cases is very similar to that of \cite{PraPraPodSva21}.

\subsection{\textbf{Case I}}
\label{sec_caseI}

In the case I ($p>2n+2$),  the {lowest} order  in  (\ref{KeyEq1C})  appears  on the {left hand} side ($\Delta^{2n-2}$). It follows that
\begin{equation}
n(n+1)=0 
\label{KeyEq1CaseI}
\end{equation}
and we are left with 
 two cases ${n=0}$ and ${n=-1}$.
 
 Taking the leading orders in each term of Eq.  (\ref{KeyEq3C}) we obtain 
\be
\big[6n(n+p-1)+p(p-1)\big]c_0\,\Delta^{n+p-2}+\cdots+2\,\Delta^{n}+\cdots -4\Lambda a_0^2\Delta^{3n}+\cdots=0\,.
\label{KeyEq3CaseI}
\ee

For ${n=0}$, ${p-2>2n=0}$ and these powers are ${\Delta^{p-2}}$,
 ${\Delta^{0}}$, and  ${\Delta^{0}}$, respectively.
 
The lowest order ${(2-4\Lambda a_0^2)\Delta^{0}}$ implies
\be
2\Lambda a_0^2=1\,.
\ee

From Eq. \eqref{KeyEq2C} we get
\be
1=\Lambda\left(2\k q^2 +\frac{8k}{3}\right)\,.
\ee
Thus, for this case, necessarily $\Lambda\not=0$.

For ${n=-1}$,  Eq. \eqref{KeyEq3CaseI} reduces to
\begin{equation}
(p-3)(p-4)c_0\,\Delta^{p-3}+\cdots +2\,\Delta^{-1}+\cdots
-4\Lambda a_0^2\Delta^{-3}+\cdots=0
\,.
\label{KeyEq3CaseIn=-1}
\end{equation}
This implies $\Lambda=0$, ${p=2}$, and  ${c_0=-1}$.

Thus, only two classes, given in Table \ref{tablecaseI} are compatible with case I.

\begin{table}[h]
		\begin{center}
				\begin{tabular}{|c||l l|}
					\hline
$[n,p]$	& constraints & \\ 
	\hline\hline
     $[-1,2]$ &  $\Lambda=0$ & $c_0=-1$  \\
 $ [0,p>2]$ & $ 1=\Lambda\left(2\k q^2 +\frac{8k}{3}\right)$ & $a_0^2=\frac{1}{2\Lambda}$ \\
    \hline
				\end{tabular} \\[2mm]
					\caption{Classes compatible with case I.}
				\label{tablecaseI}
		\end{center}
\end{table}

\subsection{\textbf{Case II}}

In the case II (${p<2n+2}$), the leading order  in (\ref{KeyEq1C}) is  $\Delta^{p-4}$  and thus from  (\ref{KeyEq1C})
\be
p(p-1)(p-2)(p-3)=0 \,.
\label{KeyEq1CaseII}
\ee 
Let us study the  four cases ${p=0}$, ${p=1}$, ${p=2}$, and ${p=3}$ separately.
 
\begin{itemize}
	\item $p=0$ \\
	In this case, the leading order of  \eqref{KeyEq3C}, $\Delta^{n-2}$, gives $n(n-1)=0$ and we are left with two cases, $n=0$, and $n=1$.
	\item $p=1$ \\
	In this case, the leading order of  \eqref{KeyEq3C}, $\Delta^{n-1}$, leads to one class $n=0$.
	\item $p=2$ \\
	In this case, two terms in Eq. \eqref{KeyEq3C} contribute to the leading order $\Delta^{n}$. The vanishing of the leading order leads to a constraint
	$(3n^2+3n+1)c_0=-1$.
	\item $p=3$ \\
	In this case, the leading order $\Delta^{n}$ cannot be set to zero, and thus this case is not compatible with \eqref{KeyEq3C}.
\end{itemize} 

In the $p=2$ case, in addition, the leading order of  \eqref{KeyEq2C},  $\Delta^{0}$, gives $2k(c_0^2 -1)-\frac{3\k q^2}{2}=0$.
 
Thus, four classes are compatible with case II, see Table \ref{tablecaseII}.

\begin{table}[h]
		\begin{center}
				\begin{tabular}{|c||l l|}
					\hline
$[n,p]$	& constraints & \\ 
	\hline\hline
      $[0,1]$ & &      \\
	$[0,0]$& &\\
    $[1,0]$& & \\
    $[n>0,2] $& $c_0^2 =\frac{3\k q^2}{4k}+1$, & $c_0=-\frac{1}{3n^2+3n+1}$\\
    \hline
				\end{tabular} \\[2mm]
					\caption{Classes compatible with case II.}
				\label{tablecaseII}
		\end{center}
\end{table}

\subsection{\textbf{Case III}}
\label{sec_caseIII}

In the case III (${p=2n+2}$), the {lowest} order $\Delta^{p-4}$  in (\ref{KeyEq1C}) implies
\be
p(p-2)\big[3a_0^ 2+4kc_0(p-1)(p-3)\big]=0\,.
\label{KeyEq1CaseIII}
\ee

Thus, there are three subcases
$[n,p]=[0,2]$,  $[n,p]=[-1,0]$, and the third case constrained by \\ ${3a_0^2=-4kc_0 (p-1)(p-3)}$, $ {p\not= 0,1,2,3}$, or equivalently 
 ${3a_0^2=-4kc_0(4n^2-1)}$. 

The leading orders of individual terms in \eqref{KeyEq3C} read 
\bea
(11n^2+6n+1) c_0\,\Delta^{3n}  +\cdots \rovno -\, \Delta^n+\cdots +2\Lambda a_0^2\Delta^{3n}\cdots \,. \label{eqtr00omegIII}
\eea
Not that this equation requires $n \leq 0$.

In the first two subcases of the case III, this gives
\begin{align}
& [n,p]=[-1,0]:\hspace{8mm} c_0=\tfrac{\Lambda}{3}a_0^2\,,\label{contrp=2c0IIIa}\\
& [n,p]=[0,2]:\hspace{8mm}  c_0=2\Lambda a_0^2-1\,.\label{contrp=2c0IIIb}
\end{align}

In the third case, the leading terms in  \eqref{KeyEq3C} read
\be
 3a_0^2=4kc_0(1-4n^2): \hspace{2mm} (11n^2+6n+1)c_0+\cdots=2\Lambda a_0^2+\cdots \,, \label{contrp=2c0IIIc}
\ee
which implies
\be
\Lambda =\frac{3}{8k}\frac{11n^2+6n+1}{1-4n^2}\quad\Rightarrow\quad c_0=\frac{3}{4k}\frac{a_0^2}{1-4n^2}\,.\label{Lc0-n,2n+2}
\ee

In the case $[0,2]$,  an additional condition follows from (\ref{KeyEq2C})  
\be
a_0^2(1-\Lambda a_0^2)(3-8k\L) =\frac{3\k q^2}{2}\,.
\ee

The three classes compatible with the case III are summarized in Table \ref{tablecaseIII}.

\begin{table}[h]
		\begin{center}
				\begin{tabular}{|c||l l|}
					\hline
$[n,p]$	& constraints & \\ 
	\hline\hline
              $ [-1,0]$ &
$c_0=\frac{\Lambda}{3}a_0^2$ &  \\
 $[0,2]$ &
$a_0^2(1-\Lambda a_0^2)(3-8k\L) =\frac{3\k q^2}{2}$, & $c_0=2\Lambda a_0^2-1$\\
$ [n<0,2n+2]$ & $\Lambda =\frac{3}{8k}\frac{11n^2+6n+1}{1-4n^2}$, & 
$c_0=\frac{3}{4k}\frac{a_0^2}{1-4n^2}$ \\
					\hline
				\end{tabular} \\[2mm]
					\caption{Classes compatible with the case III.}
				\label{tablecaseIII}
		\end{center}
\end{table}

The list of all allowed classes 	$[n,p]$  derived in Secs. \ref{sec_caseI}--\ref{sec_caseIII} is  given in Table \ref{tbl1}.

	\newpage

\section{Fine-tuned parameters}

\begin{footnotesize}
\begin{longtable}{|c|c|c|c|c|}
    \hline
    $a_0$ & $c_0$ & $q$ & $\Lambda$ & $b$ \\[0.5mm]
    \hline\hline
    $1$ & $-1$ & $0$ & $-9/10$ & $0$ \\[1mm]
    $1$ & $-1$ & $0$ & $-8/10$ & $0$ \\[1mm]
    $1$ & $-1$ & $0$ & $-7/10$ & $0$ \\[1mm]
    $1$ & $-1$ & $0$ & $-6/10$ & $0$ \\[1mm]
    $1$ & $-1$ & $0$ & $-5/10$ & $0$ \\[1mm]
    $1$ & $-1$ & $0$ & $-4/10$ & $0$ \\[1mm]
    $1$ & $-1$ & $0$ & $-3/10$ & $0$ \\[1mm]
    $1$ & $-1$ & $0$ & $-2/10$ & $0$ \\[1mm]
    $1$ & $-1$ & $0$ & $-1/10$ & $0$ \\[1mm]
    $1$ & $-1$ & $0$ & $0$ & $0$ \\[1mm]
    $1$ & $-1$ & $0$ & $1/10$ & $0$ \\[1mm]
    $1$ & $-1$ & $0$ & $2/10$ & $0$ \\[1mm]
    $1$ & $-1$ & $0$ & $3/10$ & $0$ \\[1mm]
    $1$ & $-1$ & $0$ & $4/10$ & $0$ \\[1mm]
    $1$ & $-1$ & $0$ & $5/10$ & $0$ \\[1mm]
    $1$ & $-1$ & $0$ & $6/10$ & $0$ \\[1mm]
    $1$ & $-1$ & $0$ & $7/10$ & $0$ \\[1mm]
    $1$ & $-1$ & $0$ & $8/10$ & $0$ \\[1mm]
    $1$ & $-1$ & $0$ & $9/10$ & $0$ \\[1mm]
    $1$ & $-1$ & $0$ & $11/10$ & $0$ \\[1mm]
    $1$ & $-1$ & $0$ & $12/10$ & $0$ \\[1mm]
    $1$ & $-1$ & $0$ & $13/10$ & $0$ \\[1mm]
    $1$ & $-1$ & $0$ & $14/10$ & $0$ \\[1mm]
    $1$ & $-1$ & $0$ & $15/10$ & $0$ \\[1mm]
    $1$ & $-1$ & $0$ & $16/10$ & $0$ \\[1mm]
    \hline
    $1$ & $-1$ & $0$ & $-0.9$ & $0.52$ \\[1mm]
    $1$ & $-1$ & $0$ & $-0.8$ & $0.51$ \\[1mm]
    $1$ & $-1$ & $0$ & $-0.7$ & $0.5$ \\[1mm]
    $1$ & $-1$ & $0$ & $-0.6$ & $0.49$ \\[1mm]
    $1$ & $-1$ & $0$ & $-0.5$ & $0.47$ \\[1mm]
    $1$ & $-1$ & $0$ & $-0.4$ & $0.453$ \\[1mm]
    $1$ & $-1$ & $0$ & $-0.3$ & $0.4335$ \\[1mm]
    $1$ & $-1$ & $0$ & $-0.2$ & $0.41245$ \\[1mm]
    $1$ & $-1$ & $0$ & $-0.1$ & $0.389206$ \\[1mm]
    $1$ & $-1$ & $0$ & $0$ & $0.3633018769168$ \\[1mm]
    $1$ & $-1$ & $0$ & $0.1$ & $0.33400590906606$ \\[1mm]
    $1$ & $-1$ & $0$ & $0.2$ & $0.300119$ \\[1mm]
    $1$ & $-1$ & $0$ & $0.3$ & $0.259664$ \\[1mm]
    $1$ & $-1$ & $0$ & $0.4$ & $0.209761$ \\[1mm]
    $1$ & $-1$ & $0$ & $1/2$ & $1460455736505795379977859245805530483862308/\
10^{43}$ \\[1mm]
    $1$ & $-1$ & $0$ & $0.6$ & $0.0602129$ \\[1mm]
    $1$ & $-1$ & $0$ & $0.64452367166149$ & $0.01$ \\[1mm]
    $1$ & $-1$ & $0$ & $0.65982915464606006$ & $-0.01$ \\[1mm]
    $1$ & $-1$ & $0$ & $0.7$ & $-0.07326$ \\[1mm]
    $1$ & $-1$ & $0$ & $73/100$ & $-140875755240252464448988834/10^{27}$ \\[1mm]
    \hline 
    \caption{Parameters of [0,1] solutions fine-tuned to various degrees towards physically-relevant asymptotic behaviours. Common parameters are $r_h=-1$, $k=1/2$ and $\kappa'=1$. Each set of parameters corresponds to a point on Fig. \ref{fig:q0Lambda}.}
    \label{tbl:tuned1}
\end{longtable}

\begin{longtable}{|c|c|c|c|c|}
    \hline
    $a_0$ & $c_0$ & $q$ & $\Lambda$ & $b$ \\[0.5mm]
    \hline\hline
    $1$ & $-1$ & $0.3$ & $-0.7$ & $-0.09$ \\[1mm]
    $1$ & $-1$ & $0.3$ & $-0.6$ & $-0.095$ \\[1mm]
    $1$ & $-1$ & $0.3$ & $-0.5$ & $-0.1$ \\[1mm]
    $1$ & $-1$ & $0.3$ & $-0.4$ & $-0.104$ \\[1mm]
    $1$ & $-1$ & $0.3$ & $-0.3$ & $-0.1078$ \\[1mm]
    $1$ & $-1$ & $0.3$ & $-0.2$ & $-0.11264$ \\[1mm]
    $1$ & $-1$ & $0.3$ & $-0.1$ & $-0.11836405$ \\[1mm]
    $1$ & $-1$ & $0.3$ & $0$ & $-0.1252725399$ \\[1mm]
    $1$ & $-1$ & $0.3$ & $0.01$ & $-0.12604402$ \\[1mm]
    $1$ & $-1$ & $0.3$ & $0.1$ & $-0.1338182$ \\[1mm]
    $1$ & $-1$ & $0.3$ & $0.2$ & $-0.144751$ \\[1mm]
    $1$ & $-1$ & $0.3$ & $0.3$ & $-0.15941$ \\[1mm]
    $1$ & $-1$ & $0.3$ & $0.4$ & $-0.18024$ \\[1mm]
    $1$ & $-1$ & $0.3$ & $0.5$ & $-0.21242$ \\[1mm]
    $1$ & $-1$ & $0.3$ & $0.6$ & $-0.27124$ \\[1mm]
    \hline
    $1$ & $-1$ & $0.3$ & $-0.5$ & $0.573$ \\[1mm]
    $1$ & $-1$ & $0.3$ & $-0.4$ & $0.557$ \\[1mm]
    $1$ & $-1$ & $0.3$ & $-0.3$ & $0.5418$ \\[1mm]
    $1$ & $-1$ & $0.3$ & $-0.2$ & $0.525355$ \\[1mm]
    $1$ & $-1$ & $0.3$ & $-0.1$ & $0.50759248$ \\[1mm]
    $1$ & $-1$ & $0.3$ & $0$ & $0.488250191$ \\[1mm]
    $1$ & $-1$ & $0.3$ & $0.01$ & $0.4862173$ \\[1mm]
    $1$ & $-1$ & $0.3$ & $0.1$ & $0.46698977031875791$ \\[1mm]
    $1$ & $-1$ & $0.3$ & $0.2$ & $0.443256$ \\[1mm]
    $1$ & $-1$ & $0.3$ & $0.3$ & $0.416188$ \\[1mm]
    $1$ & $-1$ & $0.3$ & $0.4$ & $0.38489$ \\[1mm]
    $1$ & $-1$ & $0.3$ & $0.5$ & $0.348684$ \\[1mm]
    $1$ & $-1$ & $0.3$ & $0.6$ & $0.30744$ \\[1mm]
    $1$ & $-1$ & $0.3$ & $0.7$ & $0.26189$ \\[1mm]
    $1$ & $-1$ & $0.3$ & $0.8$ & $0.214224$ \\[1mm]
    $1$ & $-1$ & $0.3$ & $0.9$ & $0.16831$ \\[1mm]
    $1$ & $-1$ & $0.3$ & $1$ & $0.12830671$ \\[1mm]
    $1$ & $-1$ & $0.3$ & $1.2$ & $0.0723224$ \\[1mm]
    $1$ & $-1$ & $0.3$ & $1.3$ & $0.0543894416$ \\[1mm]
    $1$ & $-1$ & $0.3$ & $1.4$ & $0.04096022$ \\[1mm]
    $1$ & $-1$ & $3/10$ & $3/2$ & $3073799795990587885810658705398202090271506172332/10^{50}$ \\[1mm]
    $1$ & $-1$ & $0.3$ & $1.6$ & $0.0228061$ \\[1mm]
    $1$ & $-1$ & $0.3$ & $1.7$ & $0.016533570014290416$ \\[1mm]
    \hline 
    \caption{Parameters of [0,1] solutions fine-tuned to various degrees towards physically-relevant asymptotic behaviours. Common parameters are $r_h=-1$, $k=1/2$ and $\kappa'=1$. Each set of parameters corresponds to a point on Fig. \ref{fig:q03Lambda}.}
    \label{tbl:tuned2}
\end{longtable}
\end{footnotesize}

\bibliographystyle{JHEP}

\bibliography{bibl,biblV2025}

\end{document}